\documentclass[8.5pt,twoside,twocolumn]{article}
\usepackage[super,sort&compress,comma]{natbib} 
\usepackage{mhchem}
\usepackage{times,mathptmx}
\usepackage{sectsty}
\usepackage{balance} 

\usepackage{color}
\usepackage[dvipsnames]{xcolor}

\usepackage{graphicx} 
\usepackage{lastpage}
\usepackage[format=plain,justification=raggedright,singlelinecheck=false,font=small,labelfont=bf,labelsep=space]{caption} 
\usepackage{fancyhdr}
\begin{document}

\thispagestyle{plain}
\fancypagestyle{plain}{
\fancyhead[L]{\includegraphics[height=8pt]{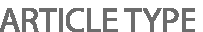}}
\fancyhead[C]{\hspace{-1cm}\includegraphics[height=20pt]{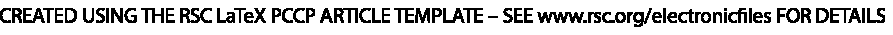}}
\fancyhead[R]{\includegraphics[height=10pt]{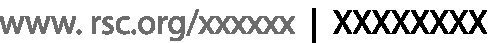}\vspace{-0.2cm}}
\renewcommand{\headrulewidth}{1pt}}
\renewcommand{\thefootnote}{\fnsymbol{footnote}}
\renewcommand\footnoterule{\vspace*{1pt}%
\hrule width 3.4in height 0.4pt \vspace*{5pt}} 
\setcounter{secnumdepth}{5}

\makeatletter 
\def\subsubsection{\@startsection{subsubsection}{3}{10pt}{-1.25ex plus -1ex minus -.1ex}{0ex plus 0ex}{\normalsize\bf}} 
\def\paragraph{\@startsection{paragraph}{4}{10pt}{-1.25ex plus -1ex minus -.1ex}{0ex plus 0ex}{\normalsize\textit}} 
\renewcommand\@biblabel[1]{#1}            
\renewcommand\@makefntext[1]%
{\noindent\makebox[0pt][r]{\@thefnmark\,}#1}
\makeatother 
\renewcommand{\figurename}{\small{Fig.}~}
\sectionfont{\large}
\subsectionfont{\normalsize} 

\fancyfoot{}
\fancyfoot[LO,RE]{\vspace{-7pt}\includegraphics[height=9pt]{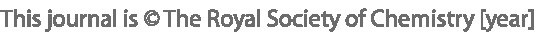}}
\fancyfoot[CO]{\vspace{-7.2pt}\hspace{12.2cm}\includegraphics{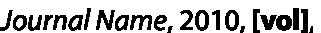}}
\fancyfoot[CE]{\vspace{-7.5pt}\hspace{-13.5cm}\includegraphics{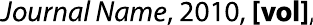}}
\fancyfoot[RO]{\footnotesize{\sffamily{1--\pageref{LastPage} ~\textbar  \hspace{2pt}\thepage}}}
\fancyfoot[LE]{\footnotesize{\sffamily{\thepage~\textbar\hspace{3.45cm} 1--\pageref{LastPage}}}}
\fancyhead{}
\renewcommand{\headrulewidth}{1pt} 
\renewcommand{\footrulewidth}{1pt}
\setlength{\arrayrulewidth}{1pt}
\setlength{\columnsep}{6.5mm}
\setlength\bibsep{1pt}

\newcommand{\mum}{~$\mu$m }
\newcommand{\mups}{$~\mu$m$\cdot$s$^{-1}$}

\twocolumn[
  \begin{@twocolumnfalse}
\noindent\LARGE{\textbf{Wall friction and Janssen effect in the solidification of suspensions
}}
\vspace{0.6cm}

\noindent\large{\textbf{Brice Saint-Michel\textit{$^{a,c}$},  Marc Georgelin\textit{$^{a}$}, Sylvain Deville\textit{$^{b}$} and Alain Pocheau*\textit{$^{a}$}}}\vspace{0.5cm}

\noindent\textit{\small{\textbf{Received 1th August 2018, Accepted 23rd October 2018\newline
First published on the web 25th October 2018}}}

\noindent \textbf{\small{DOI: 10.1039/C8SM01572D}}
\vspace{0.6cm}

\noindent \normalsize{We address the mechanical effect of rigid boundaries on freezing suspensions.
For this we perform the directional solidification of monodispersed suspensions in thin samples and we document the thickness $h$ of the dense particle layer that builds up at the solidification front.
We evidence a change of regime in the evolution of $h$ with the solidification velocity $V$ with, at large velocity, an inverse proportionality and, at low velocity, a much weaker trend.
By modelling the force balance in the critical state for particle trapping and the dissipation phenomena in the whole layer,  we link the former evolution to viscous dissipation and the latter evolution to solid friction at the rigid sample plates.
Solid friction is shown to induce an analog of the Janssen effect on the whole layer.
We determine its dependence on the friction coefficient between particles and plates, on the Janssen's redirection coefficient in the particle layer, and on the sample depth.
Fits of the resulting relationship to data confirm its relevance at all sample depths and provide quantitative determinations of the main parameters, especially the Janssen's characteristic length and the transition thickness $h$ between the above regimes.
Altogether, this study thus clarifies the mechanical implication of boundaries on freezing suspensions and, on a general viewpoint, provides a bridge between the issues of freezing suspensions and of granular materials.
}
\vspace{0.5cm}
 \end{@twocolumnfalse}
  ]



\footnotetext{\textit{$^{a}$~Aix Marseille Univ, CNRS, Centrale Marseille, IRPHE, Marseille, France E-mail: alain.pocheau@univ-amu.fr}}
\textcolor{black}{\footnotetext{\textit{$^{b}$~Laboratoire de Synth\`ese et Fonctionnalisation des C\'eramiques, UMR3080 CNRS/Saint-Gobain CREE, Saint-Gobain Research Provence, Cavaillon, France}}}
\footnotetext{\textit{$^{c}$~Now at: Department of Chemical Engineering, Imperial College London, London SW7 2AZ , United Kingdom}}



\section{Introduction}

The solidification of a suspension (i.e. of a two phase mixture involving solid particles dispersed in a fluid) arises in a number of situations either natural as the freezing of soils \cite{Corte1962,Zhu2000,Rempel2004,Dash2006,Peppin2013,Saruya2013,Anderson2014,Saruya2014} or man-made as in the food industry \cite{Velez-Ruiz2007}, the casting of particle rich alloys \cite{Stefanescu1988,Asthana1993,Liu2008} or the making of bio-inspired composite materials \cite{Deville2007a}.
Its physics involves phenomena referring either to solidification, to suspension, or to the interaction between a solidification front and the suspension particles.
The latter phenomenon gives rise, by van der Waals or electrostatic interactions,  to a  thermomolecular force between front and particles that is usually repulsive.
It then results in the formation of a dense layer of particles that is pushed by the advancing solidification front and in which phenomena pertaining to the physics of suspensions may occur.
However, 
\textcolor{black}{the specific effects}
induced by an assembly of particles have been poorly addressed to date in the context of suspension freezing.
Here, we wish to study them by addressing the effects of boundaries on the compacted particle layer formed ahead of the front.

Boundaries will be provided by the sample plates that sandwich the suspension.
Attention will be focused on the steady thickness reached by the compacted particle layer when particles succeed in being trapped by the front.
In particular, its evolution with respect to the solidification velocity and the sample depth will be documented at several sample depths.
Two regimes will be found : one, dominated by viscous friction of the fluid flowing across the dense layer to feed solidification ; the other, dominated by solid friction between particles and plates.

Whereas the viscous regime has been recently studied in detail first at a single \cite{Saint-Michel2017}, then at multiple sample depths \cite{Saint-MichelSubmitted}, the solid friction regime has not, to the best of  our knowledge, been documented yet.
It will be the objective of this study to clarify it.

The combined mechanical model of particle layer and particle trapping proposed in 
ref.
\cite{Saint-Michel2017,Saint-MichelSubmitted} will be modified to include Coulomb frictional contacts at the sample plates and the resulting stress redistribution in the particle layer.
The new model will enable to recover the main features of both regimes as well as the transition between them.
In particular, a cumulative effect of friction will become prominent in the solid friction regime, when the compacted layer exceeds some characteristic size.
This effect is similar  to the Janssen effect in granular materials following which the apparent weight of granular columns confined in silos is bounded at a value corresponding to a characteristic height $\lambda$, thanks to cumulative friction effects at the silo's walls and to stress redistribution in the granular column \cite{Janssen1895,JanssenSperl1895,Bertho2003,Andreotti2013}.
The length $\lambda$, called the Janssen's length, will state here the transition between the above regimes in terms of layer thickness.
Its variation with the sample depth $e$ will be evidenced and found to agree with the modelling.
This will allow us to clarify  its link with mechanical coefficients such as the friction coefficient and the Janssen's redirection coefficient. 
In addition, an indirect measurement of the thermomolecular repulsive force between the solidification front and a nearby particle will be performed from the fits of data to the relationship derived from modelling.

Altogether, this study thus clarifies the effect of solid boundaries on solidifying suspensions and provides a bridge between the physics of granular suspensions and that of freezing suspensions.

\section{Experiment}
\label{Experiment}

The experiment aims at performing the solidification of suspensions in a controlled way so as to allow quantitative analyses in link with appropriate modelling.
For this, care is taken to manage the main parameters of solidification (temperature gradient, growth velocity), of suspensions (monodispersity, particle volume fraction), and of sample geometry (parallelepiped, depth) and to obtain relevant real-time visualization of the freezing suspension.

\subsection{Setup}
\label{Setup}

\begin{figure*}[t!!]
\centering
\includegraphics[width=8cm]{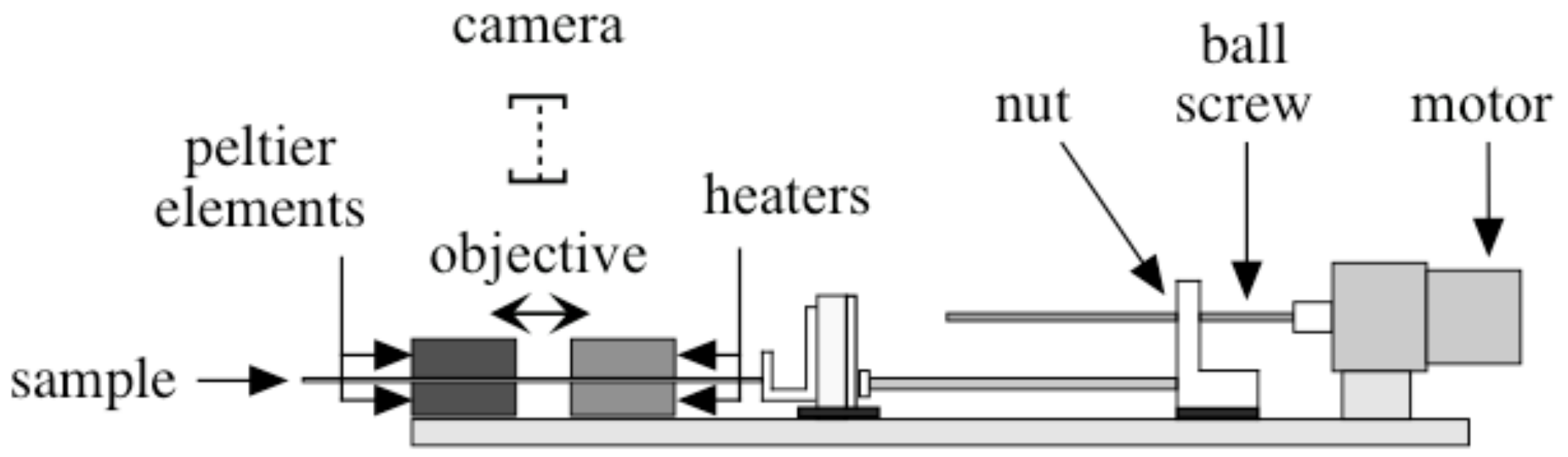}
\hspace{0.5cm}
\includegraphics[width=8cm]{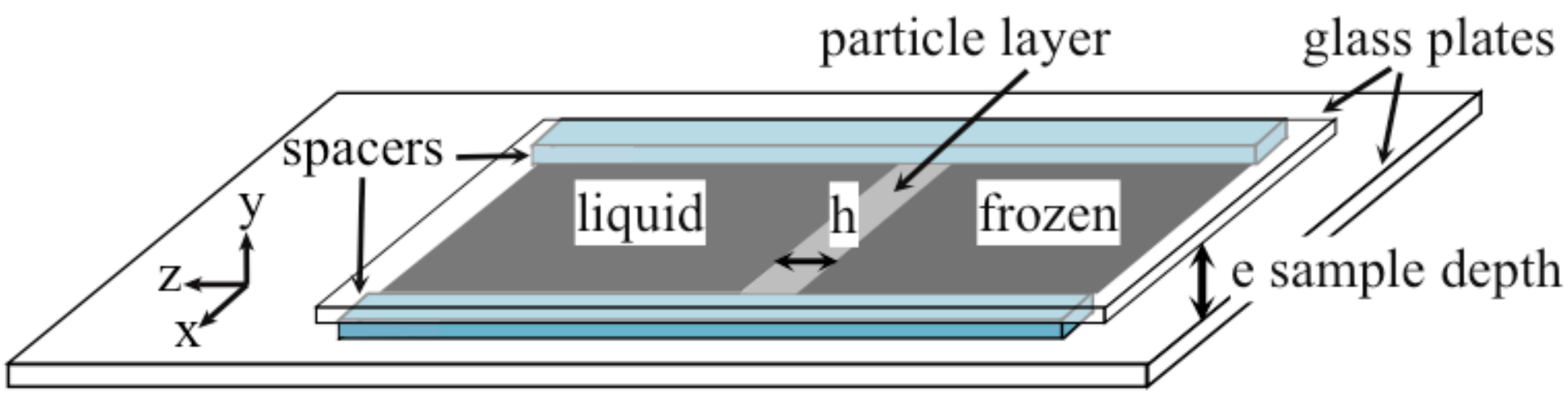}
\\
\flushleft{\hspace{5cm} \textcolor{black}{(a)} \hspace{7cm} \textcolor{black}{(b)}}
\caption{(Color online)
Sketch of the experimental setup. 
(a) A thin sample is pushed in between heaters and coolers by a micro-stepper motor coupled to a linear track.
Real-time non-invasive visualization of the solidification interface is provided by an optical access through which an objective allows the formation of a magnified image on a camera.
(b) Samples containing the suspension are made of two glass plates separated by calibrated spacers which enable the sample depth $e$ to be varied.
Suspensions are filled in by capillarity into a large domain, $100$ mm long and $45$ mm wide, that is sealed before solidification.
In a freezing suspension, a compacted particle layer of thickness $h$ stands ahead of the solidification front.
The axes $x$, $y$ and $z$ refer to the directions of the solidification front, the sample depth and the thermal gradient respectively.
}
\label{Set-up}
\end{figure*}

The experimental setup has been adapted from a previous one dedicated to the solidification of binary solutions \cite{Georgelin1998,Pocheau1999}.
It aims at solidifying a thin sample at a prescribed velocity in a fixed thermal gradient while allowing visualization of the vicinity of the solidification front [Fig.\ref{Set-up}(a)].

The sample consists of glass plates separated by calibrated spacers that delimit a parallelepiped space in which the suspension is introduced by capillarity [Fig.\ref{Set-up}(b)].
Its dimensions (top glass $100\times 45 \times 0.7$~mm$^3$ ; bottom glass $150\times 50 \times 0.8$~mm$^3$) have been taken long and large enough to allow a large central zone ($35$~mm at least) free of lateral boundary effects and compatible with solidification over hours.
Observations will always refer to it.
The sample depth can be changed by varying the spacer thickness.
In practice, six depths were studied, $16, 30, 50, 75, 100$ and $125 \mu$m, each in specific samples but with otherwise the same setup and the same suspensions.
Samples were sealed with cyanoacrylate and epoxy glue.

The suspensions, manufactured by Magsphere Inc., were stable over months.
They contain plain polystyrene (PS) spheres of density $1.05$ and volume fraction $\phi_0=10\%$ or $20\%$ $\pm 0.5 \%$.
Their size distribution, measured by a Coulter counter
\textcolor{black}{by the manufacturer,}
leads a mean diameter of $3.0~\mu$m with a relative standard deviation as low as $4\%$, thus resulting in quasi mono-dispersed suspension.
The solvent was water with a small amount of surfactant and less than $0.09 \%$ of sodium azide.
When solidified alone, it led a planar front to destabilize at a large growth velocity of several \mups{}, thus indicating a low concentration of additive.
Its dynamical viscosity $\mu$ was thus taken as that of water : $\mu=1.8\times 10^{-3}$ Pa.s.

The sample was pushed on a linear track (THK) by a microstepper motor (ESCAP) driving a recirculating ball screw (Transroll).
The elementary displacement on a microstep was, with $6400$ microsteps by turn and a $5$ mm screw pitch, as small as $0.8 \mu$m.
At the end of each microstep, the motor uses an electronic damping to slow down its rotation and thus prevent vibrations.
The available velocity range extends from $0.07$ to $50~\mu$m.s$^{-1}$ with relative modulations less than $3\%$.

Heaters and coolers sandwich the sample so as to induce a controlled thermal gradient in it.
They are made of copper blocks either heated by resistive sheets (Minco) or cooled by Peltier devices (Melcor) and are separated by a $10$ mm gap.
They are electronically regulated at temperatures of $\pm 20^{\circ}$C following which the melting isotherm takes place in the middle of the gap.
This facilitates visualization in the optical window that stands in between the thermal blocks (Fig.\ref{Set-up}-a) and minimize the dependence of the thermal gradient on the sample velocity $V$ \cite{Georgelin1998,Pocheau2009}.
Extra heat released by the Peltier devices or conducted from the heaters to the lateral sides of the setup were extracted by an external circulation of cryogenic fluid at $- 30^{\circ}$C.
Insulating polystyrene walls were finally placed all around the whole setup so as to provide a closed dry atmosphere and help avoiding condensation and ice formation.

Visualization was achieved by forming an image of the solidification front on a CCD camera.
Care has ben taken to avoid thermal perturbation by placing the optical devices far from the front.
For this, a home-made optical setup with a large focal length of $50$ mm has been preferred to a microscope.
A photographic lens was then placed at a distance of about a focal length from the solidification front so as to provide an enlarged image on a camera placed about a meter apart.
Despite its simplicity, an excellent image sharpness could be 
 \textcolor{black}{obtained}
since the low inclination of the optical rays enables the Gauss approximation and stigmatism to be satisfactorily fulfilled.

\begin{figure}[h!]
\includegraphics[width=8.5cm]{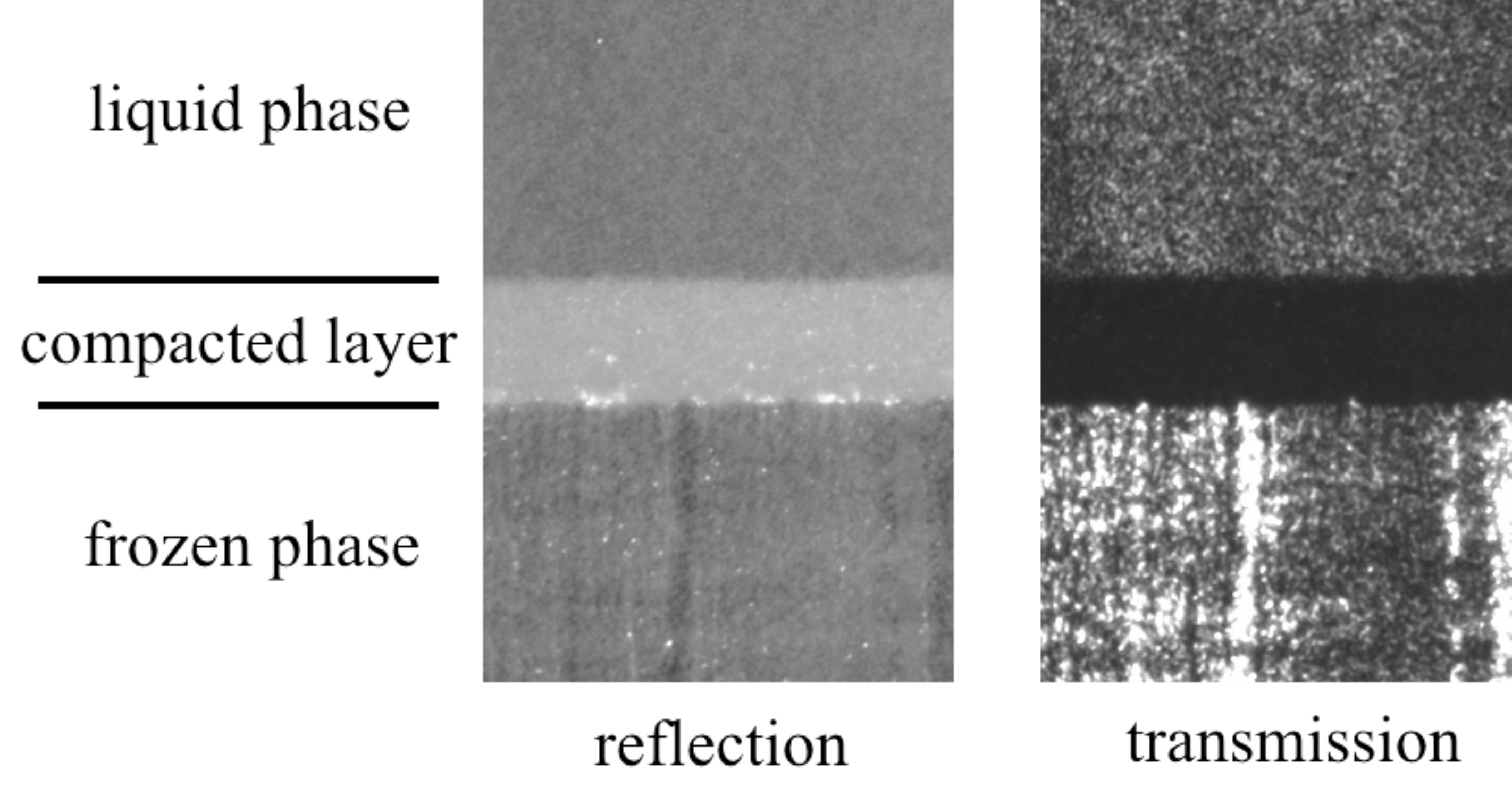}
\caption{
Images of the a compacted layer by reflection (left) and transmission (right) : $\phi_0 = 20\%$ , $e=16 \mu$m ; image width 640 $\mu$m.
They involve a top zone (the liquid phase), a bottom zone (the frozen phase) and an intermediate zone (the compacted particle layer).
Bright zones refer to large $\phi$ by reflection (left) and to low $\phi$ by transmission (right).
The change of optical methods thus reverses the zone intensity : the top and bottom zones remain grey but the intermediate zone turns from bright to dark.
We notice that, regardless of the optical method, the compacted layer displays the same extent.
This indicates that it keeps the same thickness from one plate to the other.
}
\label{ReflectionTransmission}
\end{figure}

Samples were observed either by  transmission or by reflection.
In the former case, light crossed the sample ; in the latter case, it was reflected by the particles.
In both cases, the image intensity was linked to the particle volume fraction $\phi$ : at large $\phi$, low intensity in transmission but high intensity in reflection, and conversely at small $\phi$.
Both methods reveal in figure \ref{ReflectionTransmission} the dense particle layer that forms ahead of the front as dark (transmission) or bright (reflection).
At the scale of the layer, its transition with both the suspension and the solid phase is sharp, both by reflection and transmission. 
Some modulations appear on the transverse direction parallel to the solidification interface but do not yield instabilities in the velocity regime studied here.
In practice, in the following, we mostly employed the reflection method.

Interestingly, Figure \ref{ReflectionTransmission} reveals that the extent of the compacted layer is the same whatever the optical method.
As the reflection method probes the vicinity of the plates up to one or two particle diameters and the transmission method the whole layer from one plate to the other, the fact that the same image of the particle layer is displayed except for the symmetry bright/dark indicates that the layer thickness remains constant from one plate to the other.
In particular, if the compacted layer had been significantly curved in between the plates, the image by transmission would have displayed a continuous transition between the grey suspension and the dark zone, in the domain where the compacted layer would have partially filled the space between the plates.

\begin{figure}[t!]
\centering
\includegraphics[height=5cm]{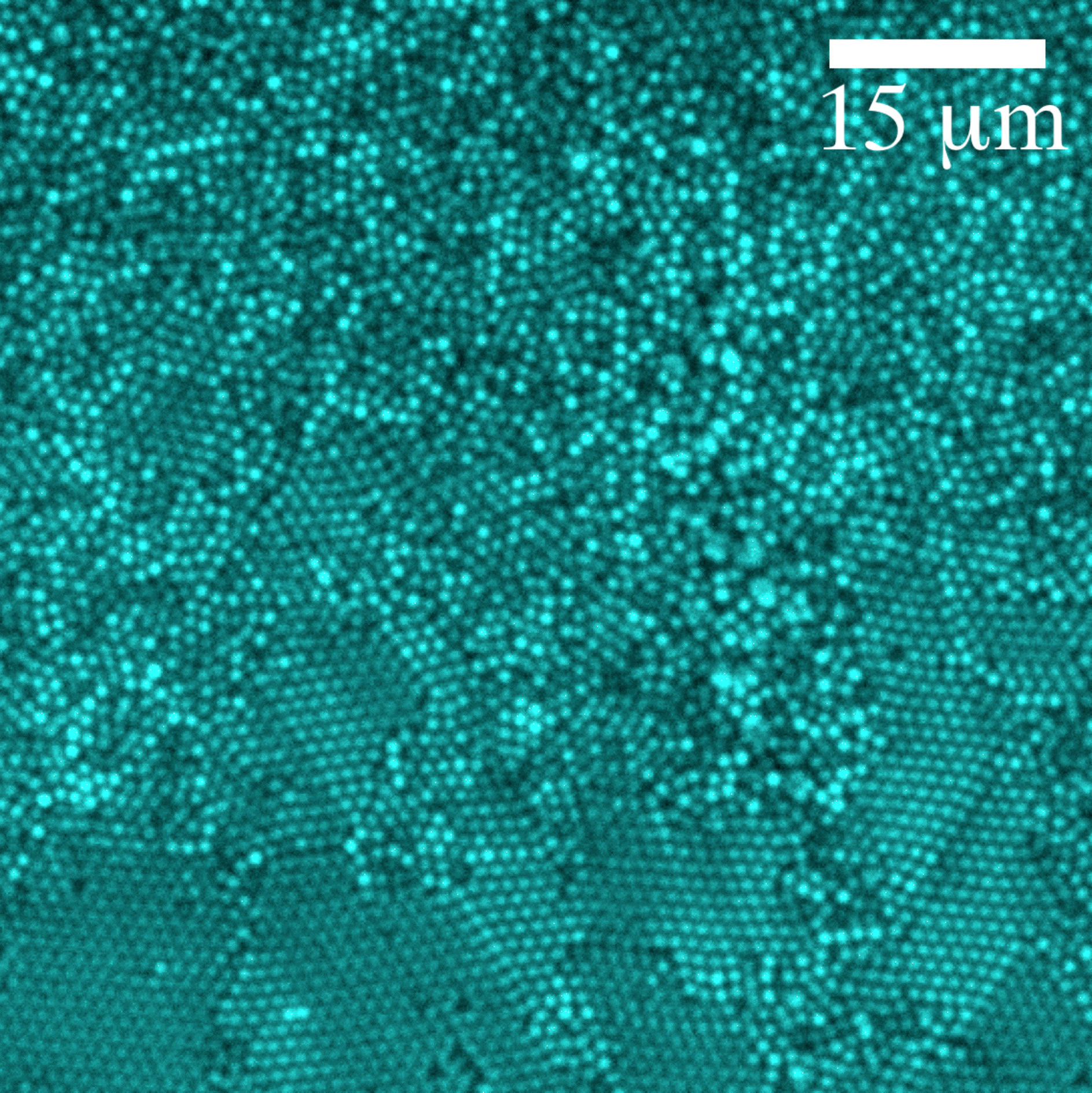}
\caption{(Color online)
Image of the compacted layer of particles at a sample plate by confocal microscopy . 
The particle diameter is $1~\mu$m. 
The image shows close packed particles filling a plane adjacent to the plate.
A particle ordering spanning at least few particles is noticeable all over the image.
It even extends to a large-scale hexagonal order at its bottom. 
}
\label{Confocal}
\end{figure}

The arrangement of particles adjacent to a plate is revealed in figure \ref{Confocal} by confocal microscopy.
Particles appear to be close packed within a plane parallel to the plate.
They show an ordering over a characteristic distance of few particles and, in the bottom of the figure, over a much larger range.
This possible long-range order corroborates the quasi mono-dispersed nature of the suspension and confirms their diameter value.
These long-range ordered patches correspond to the bright patches observed by light reflection in figure \ref{CompactLayer}.
This particle ordering is presumably triggered by the impenetrable plate which forces adjacent particles to lie in a plane.
It is reminiscent of the ordering induced by planar boundaries in monodisperse granular materials \cite{Zhang2006,Burtseva2015,Mandal2017} or of the particle crystallization evidenced near the walls in monodisperse mixtures in channel flows \cite{Mandal2017}, Couette flows \cite{Mueth2003}, or by shaking \cite{Pouliquen1997}.
It is expected to fade away with the distance to the plate and to disappear beyond a few particle diameters from the plate to leave place for a random arrangement.

As the sample is horizontal and the solidification velocities low, particles have enough time to sediment on the bottom plate before encountering the advancing compacted particle layer.
As they remained dispersed, they then redistribute 
\textcolor{black}{in between the sample plates}
and fill the compacted layer in the whole sample depth at the same rate as if the suspension had remained homogeneous.
As the essential physical mechanisms will concern the particle layer and the solidification front, they are thus unaffected by 
\textcolor{black}{this}
transient sedimentation.
Its sole observable effect then stands in a thin redistribution zone ahead of the compacted layer which rounds its transition with the suspension over a distance that we estimate to a sample depth.
As the layer thickness $h$ was measured after thresholding, this affected its measurement accuracy.

Some additional phenomena induced by particles or solidification may appear intermittently and induce fluctuations.
They refer to the compaction of some particle patches that makes them enter the front as a whole, to the stress propagation in the particle layer which is known to 
\textcolor{black}{involve}
intrinsic fluctuations \cite{Bouchaud1995}, to  grain boundaries at the solidification front that favor particle trapping ...
They are responsible 
\textcolor{black}{for}
small undulations of the particle layer (Figs.\ref {ReflectionTransmission} and \ref{CompactLayer}) and thus to fluctuations of $h$.
Their relative amplitude has been measured to be $15\%$ at most and $6\%$ in average and may
\textcolor{black}{be}
 taken as an estimation of the width of error bars.

In the remainder of the paper, we assign the $x$-axis, $y$-axis and $z$-axis to the directions of the solidification front, the sample depth and the thermal gradient respectively [Fig.\ref{Set-up}(b)].
Samples are thus pushed in the $z$ direction along which the dense particle layer develops.

\subsection{Particle layer }
\label{Particle layer }

\begin{figure}[t!]
\centering
\includegraphics[width=8.5cm]{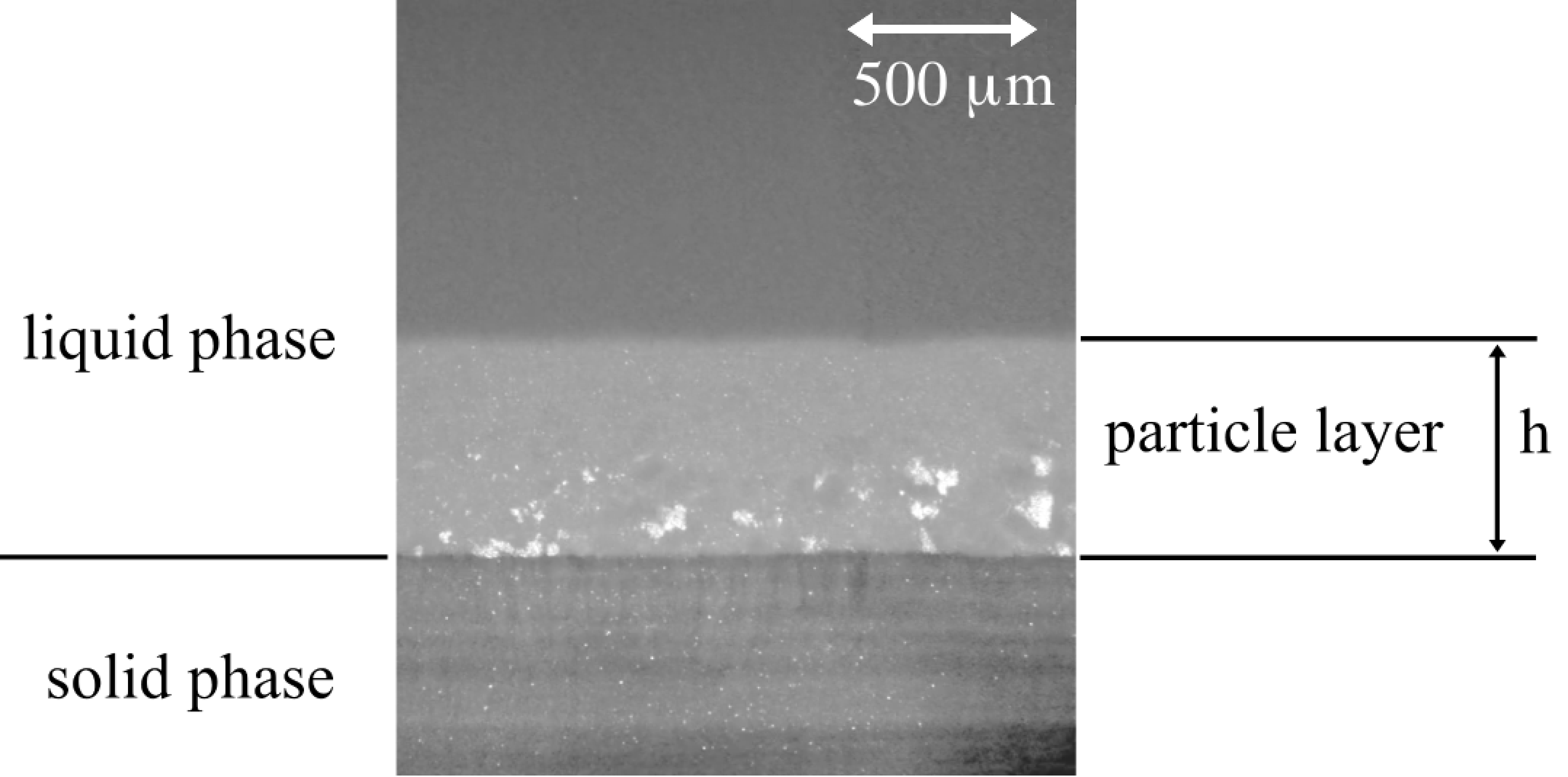}
\caption{
Compact layer of particles formed ahead of the solidification front : $\phi_0 = 20\%$ , $e=16 \mu$m, $V=1$\mups{}.
The image is obtained by reflection.
Its top corresponds to the liquid phase and its bottom to the frozen phase. 
Both appear grey as their particle volume fraction, $\phi_0$, is low.
In between them, the bright zone reveals a noticeable increase of particle density which corresponds to the compacted particle layer.
It involves bright spots which correspond to crystallized patches.
The steady state thickness reached by the particle layer beyond its build-up is labelled $h$.
}
\label{CompactLayer}
\end{figure}

At the beginning of solidification, the solidification front faces a homogeneous suspension with a particle volume fraction $\phi_0$.
However, in the velocity range of the experiment, particles are first repelled by the solidification front.
They then accumulate ahead of it in a dense layer that is visible in figure \ref{CompactLayer} as a bright zone induced by light reflection.
Meanwhile, the thickness of the layer grows at a constant rate that we have quantitatively studied in the same setup at the largest sample depth $e=125 \mu$m \cite{Saint-Michel2017}.
By applying particle conservation, this provided the average particle volume fraction $\phi_l$ of the layer, $\phi_l=0.634 \pm 0.007$, equal to the random close packing value $\phi_{rcp}=0.634$.
 This is thus consistent with a random accumulation of particles in the layer.
 
 At some thickness $h$, the dense layer stops growing and particles appear in the solid phase.
 Despite repelling forces exerted by the solidification front, particles then succeed in entering the ice phase at a rate equal to their incoming rate in the layer.
 This steady value $h$ of the layer thickness is thus linked to the repelling/trapping transition of particles by the front.
 Interestingly, it thus corresponds to a macroscopic variable linked to a micronic or submicronic phenonemon.
 In this regard, it may enable us to get insight into a small scale issue by a large scale measurement.
 We shall thus focus our attention on it from now on.
 It will appear to depend on the particle volume fraction $\phi_0$, the growth velocity $V$ and the sample depth $e$.

\begin{figure}[h!]
\centering
\includegraphics[width=6cm]{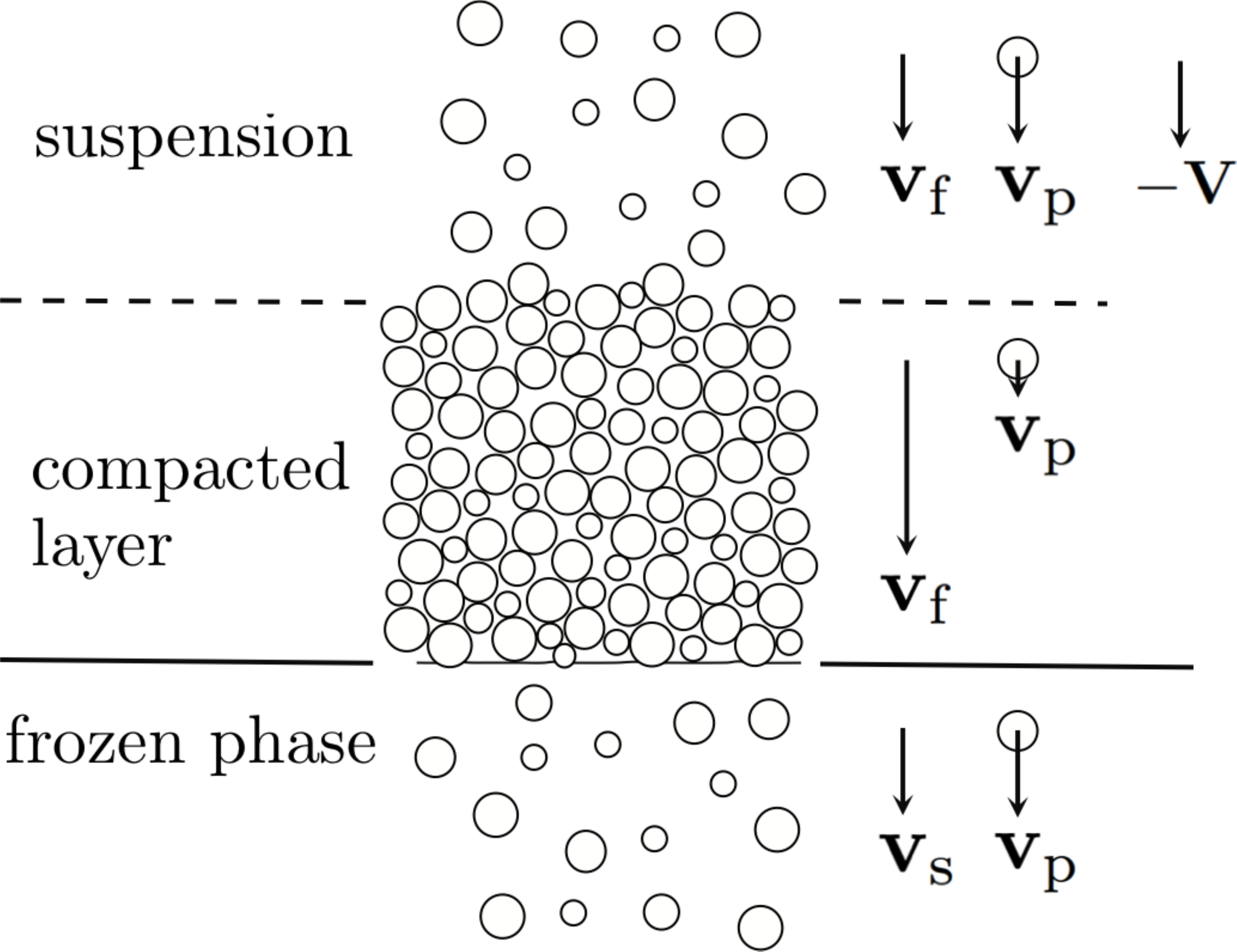}
\caption{Sketch of a cross-section of the suspension in the $z$ direction of the thermal gradient.
It displays  the suspension, the compacted layer of particles and the frozen phase.
As particles share the same radius, the apparent variation of their radii is an artifact of the section, the particles being randomly distributed in space.
The mean velocities with respect to the solidification front of the liquid fluid, the particles and the solid phase  are denoted $\mathbf{v}_{\rm f}$, $\mathbf{v}_{\rm p}$ and $\mathbf{v}_{\rm s}$ respectively.
In the suspension and in the solid phase, they all equate to the opposite $- \mathbf{V}$ of the solidification velocity. 
However, in the compacted layer of particles, $\mathbf{v}_{\rm f}$ and $\mathbf{v}_{\rm p}$ differ following the rise of particle volume fraction $\phi$.
}
\label{Suspension}
\end{figure}

 Following the build-up of the dense particle layer, the particle volume fraction has become inhomogeneous with a step from $\phi_0$ to $\phi_{rcp}$ at the 
 \textcolor{black}{frontier}
  between the suspension and the dense layer and the reverse step at the solidification front.
When the layer thickness is steady, mass conservation of fluid and particles implies a related step in their respective velocities with respect to the solidification front, $\mathbf{v}_{\rm f}= v_{\rm f} \; \mathbf{e}_z$ for the fluid and $\mathbf{v}_{\rm p} = v_{\rm p} \; \mathbf{e}_z$ for the particles (Fig.\ref{Suspension}).
 In particular, whereas both velocities are opposite to the growth velocity in both the suspension and the solid phase, $ v_{\rm f}= v_{\rm p}=-V$, they amount to $v_{\rm f} = -V (1-\phi_0)/(1-\phi_l)$ and $v_{\rm p}=-V \phi_0/\phi_l$ in the layer.
 As they differ, a viscous dissipation is generated.
 It depends on the volume flux of fluid $\mathbf{U}= U\mathbf{e}_z$ with respect to the particle matrix, $U=(1-\phi_l)(v_{\rm f}-v_{\rm p})$, i.e. :
 \begin{equation}
\label{U}
U = - V \frac{(\phi_l-\phi_0)}{\phi_l}
\end{equation}
As $\textcolor{black}{\phi_l} > \phi_0$, $U$ is negative, so that the viscous force exerted by the fluid on the particle matrix 
\textcolor{black}{is}
directed towards the front.
To avoid sign confusion,  its magnitude will be hereafter labeled $|U|$.

 The velocity $\mathbf{U}$, called the Darcy velocity, is thus more relevant to the physical process at work in the layer than the growth velocity $\mathbf{V}$.
 In particular, using it instead of $\mathbf{V}$ to report the evolution of $h$ leads to a collapse of data on master curves (see for instance Fig. \ref{G,h,U,e125})
 instead of a slight shift depending on $\phi_0$ otherwise \cite{Saint-Michel2017}.
 We shall thus adopt it in the rest of the paper.
 This will reduce the dependence of $h$ on $U$ and $e$ only.
\textcolor{black}{The}
relative fluctuations 
\textcolor{black}{of $U$}
will result from those of $V$ and of $\phi_0$ and amount to $4\%$.
 
\section{Particle layer thickness}

We report the evolution of the steady layer thickness $h$ with respect to both the magnitude $|U|$ of the Darcy velocity and the sample depth $e$.

\begin{figure*}[t!]
\centering
\includegraphics[width=5cm]{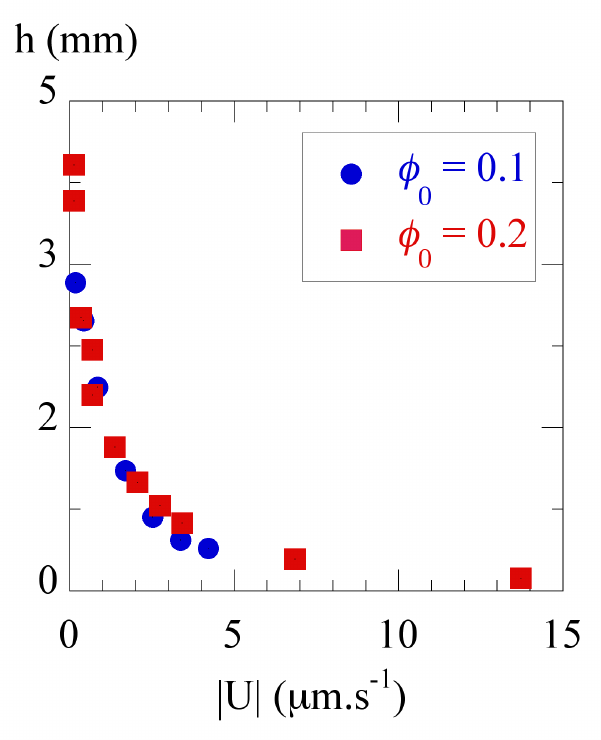}
\hspace{0.5cm}
\includegraphics[width=5cm]{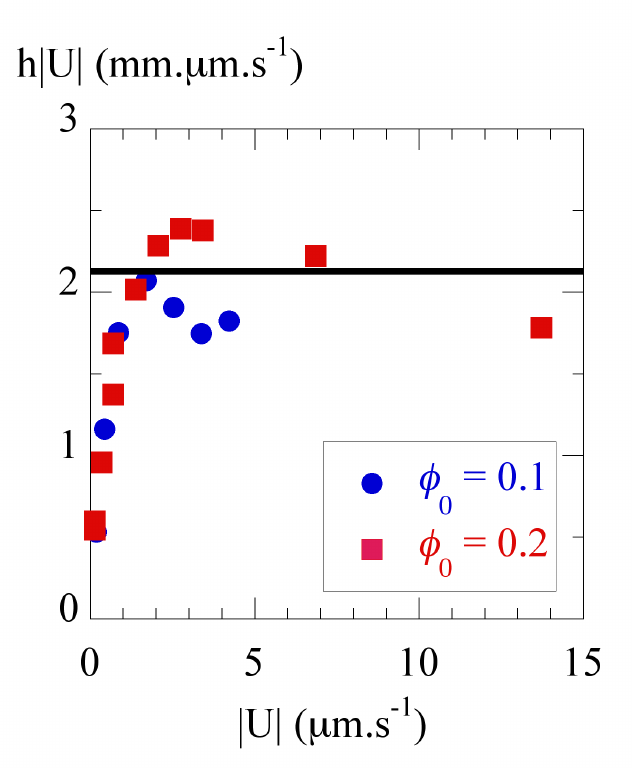}
\hspace{0.5cm}
\includegraphics[width=5cm]{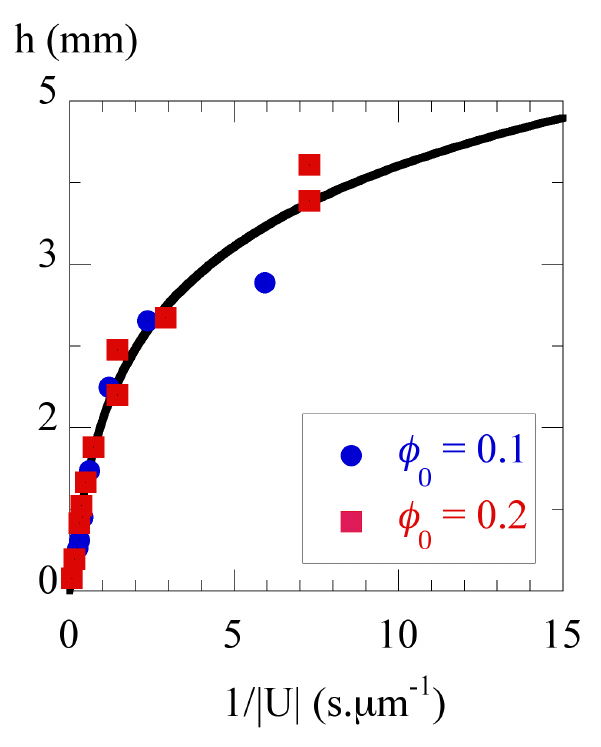}
\\
\flushleft{\hspace{3cm} (a) \hspace{5cm} (b) \hspace{5cm} (c)}
\caption{(Color online)
Evolution of the layer thickness $h$ with the Darcy velocity $|U|$ at sample depth $e=125 \mu$m.
(a) $h(|U|)$ : $h$ shows a continuous decrease with $|U|$ ;
(b) The plot of $h |U|$ as a function of $|U|$ shows that, for $|U|$ larger than about $2 \mu$m.s$^{-1}$, $h$ is inversely proportional to $|U|$. The line provides the data average of $h |U|$ for $|U|>2 \mu$m.s$^{-1}$.  ;
(c) The plot of $h$ as a function of the inverse Darcy velocity $1/|U|$ shows a linear trend up to about $0.5$ s.$\mu$m$^{-1}$, followed by a concave trend.
The curve corresponds to a fit of data to relation (\ref{h,U,dpdz}).
}
\label{G,h,U,e125}
\end{figure*}

\subsection{Evolution with the Darcy velocity $U$ at a given $e$}

We focus our attention here on $e=125 \mu$m which is representative of the evolution at any sample depth $e$.
Figure \ref{G,h,U,e125}-a shows a decrease of $h$ with the Darcy velocity $U$.
However, a finer analysis of this evolution in figures \ref{G,h,U,e125}-b and \ref{G,h,U,e125}-c reveals two different trends  :

First, the graph of the product $h |U|$ with respect to $|U|$ shows in figure \ref{G,h,U,e125}-b that, above about $|U|=2 \mu$m.s$^{-1}$, $h$ is inversely proportional to $|U|$.
This corresponds in figure \ref{G,h,U,e125}-c to the linear trend displayed at low $1/|U|$ below $0.5 \; s.\mu$m$^{-1}$.
In contrast, at small velocities, figure \ref{G,h,U,e125}-b shows that, below about $|U|=2 \mu$m.s$^{-1}$, $h$ decreases quicker than $|U|^{-1}$, as $|U|^{-2}$ at least.
This is confirmed by the concavity of the graph of figure \ref{G,h,U,e125}-c above $1/|U|\approx 1 \mu$m.s$^{-1}$, which reveals a much weaker growth of $h$ with $1/|U|$ at very low velocity.

These two trends signal the occurrence of two different physical regimes.
At large velocities, the trend $h \propto 1/|U|$ will be related in section \ref{Modelling} to a dominant viscous dissipation in the particle layer.
At low velocities, the much weaker variation of $h$ with $|U|$ will be related in the same section to a dominant dissipation provided by solid friction between the particles and the sample plates.

\subsection{Evolution with the Darcy velocity $U$ at various $e$}

When decreasing $e$ from $125 \mu$m to $16 \mu$m, figure \ref{G,h,1-U,NL} shows similar trends of $h$ with respect to $|U|$.
In particular, at all sample depths $e$, one recovers at large velocities a linear 
increase of $h$ with $1/|U|$ followed, at low velocities, by a concave evolution towards a much slower raise.
However, beyond these qualitative similarities, the graphs show quantitative differences.
In particular, the transition between the two different trends is encountered at lower $h$ and lower $1/|U|$ as $e$ decreases.
Similarly, both the slope of the linear trend $h\propto 1/|U|$ displayed at low $1/|U|$ and the value of $h$ reached at $1/|U|=15$ s.$\mu$m$^{-1}$ decrease with decreasing $e$.

The remainder of the paper will be devoted first to understand each regime and their transition, then to model the evolution of the layer thickness on the whole velocity range and, finally, to determine the relevant physical variables and their evolution with $e$.
This will yield in figure \ref{G,h,1-U,NL} the non-linear fits displayed by full lines and a location of the transition at the 
\textcolor{black}{dashed}
lines $h=\lambda$ with a transition thickness $\lambda(e)$ dependent on $e$.
As these fits satisfactorily recover data in figure \ref{G,h,1-U,NL}, we gather them in figure \ref{G,h,1_U,Alle,fts} to get an overview of the evolution of the whole data.
It shows both the similarities between the evolutions of $h(1/|U|)$ at different $e$ and their quantitative differences.

\begin{figure*}[t!!!!!]
\centering
\includegraphics[height=5.5cm]{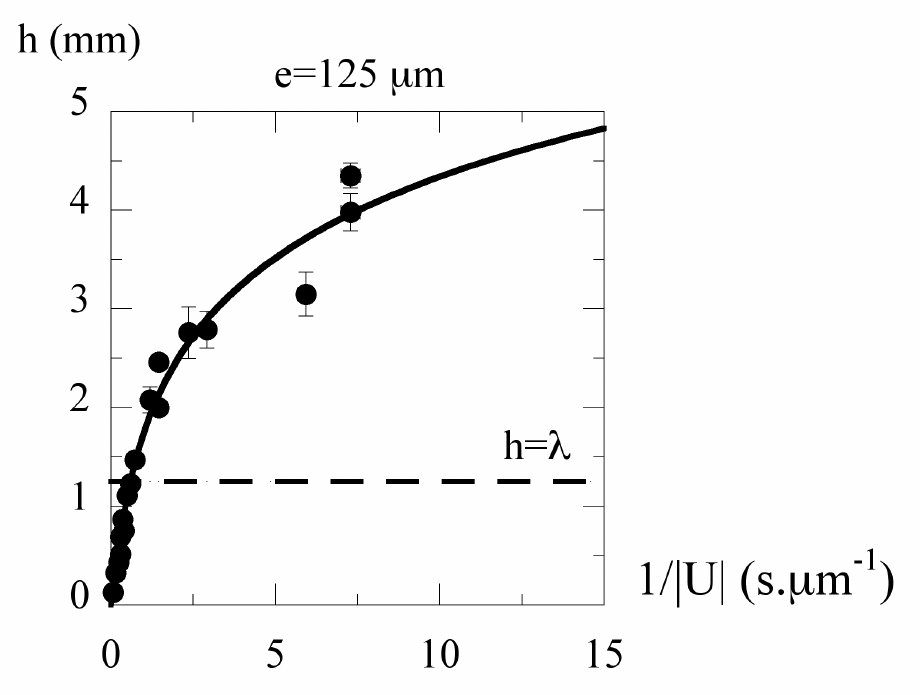}
\hspace{1.5cm}
\includegraphics[height=5.5cm]{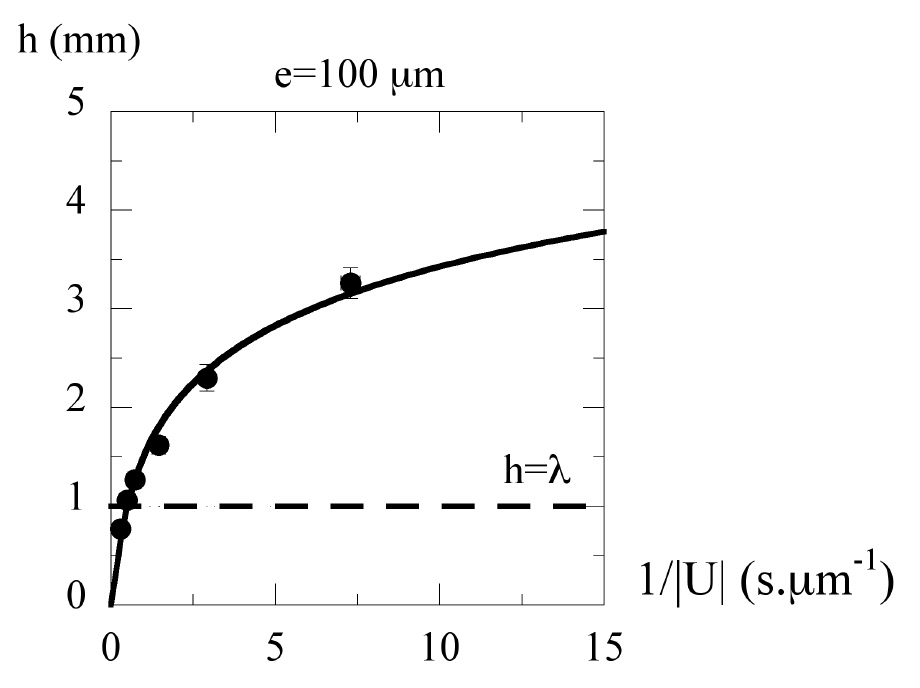}
	\flushleft{\hspace{4cm} (a) \hspace{8cm} (b)}
	\\
\centering
\includegraphics[height=5.5cm]{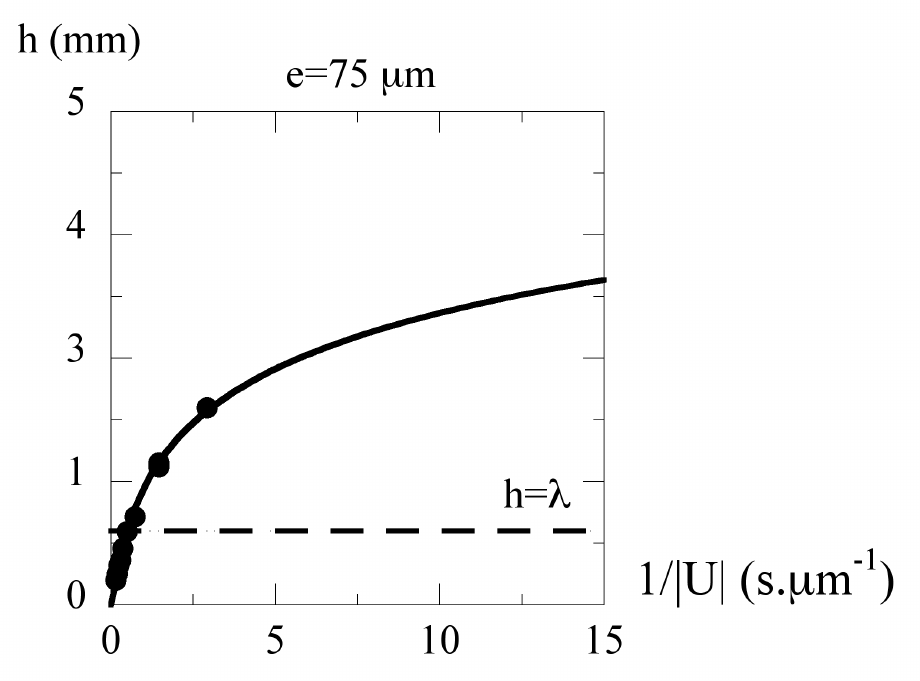}
\hspace{1cm}
\includegraphics[height=5.5cm]{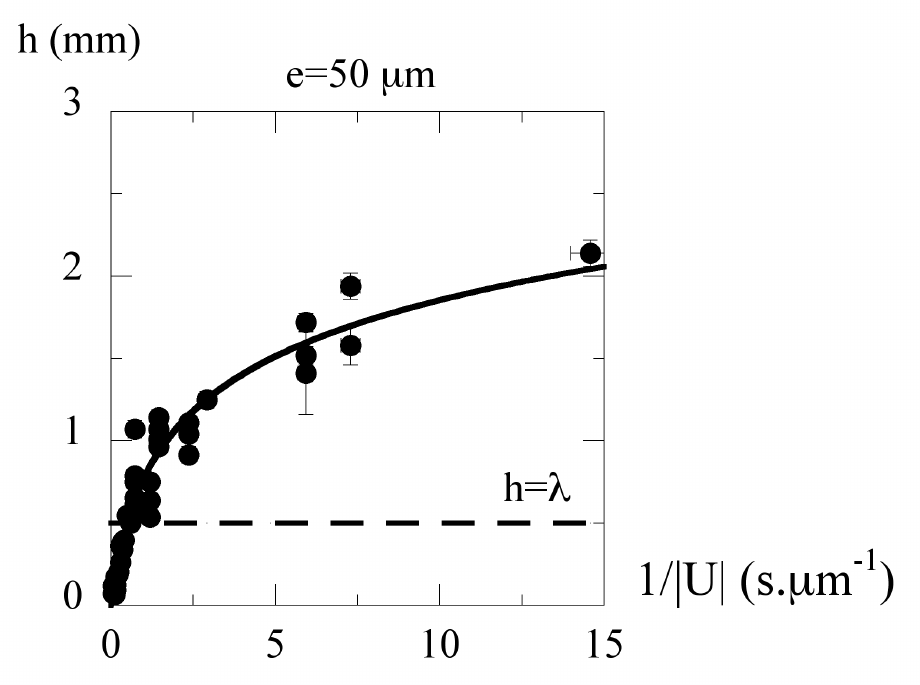}
	\flushleft{\hspace{4cm} (c) \hspace{8cm} (d)}
	\\
\centering
\includegraphics[height=5.5cm]{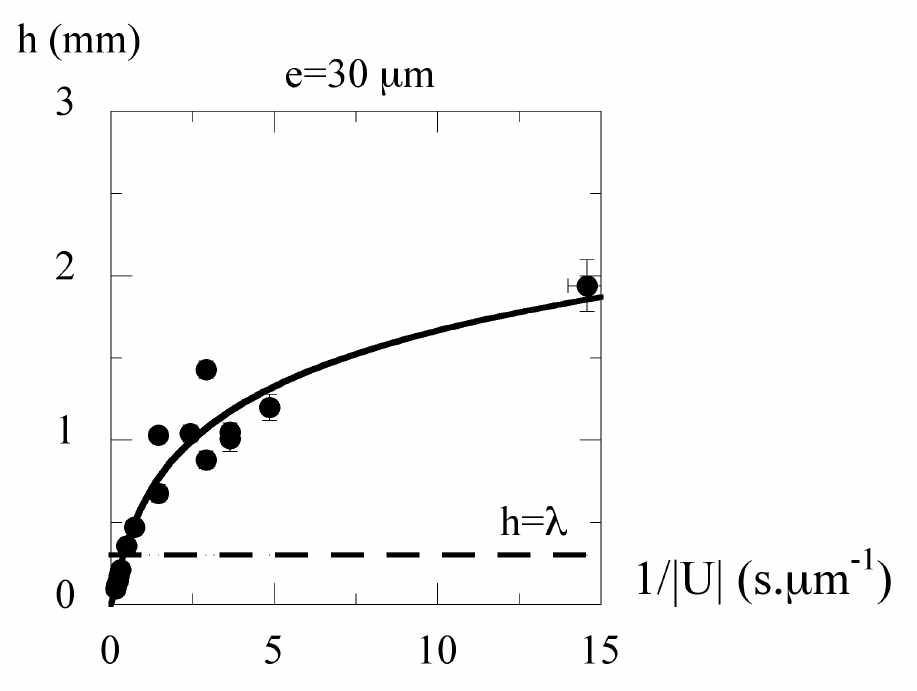}
\hspace{1cm}
\includegraphics[height=5.5cm]{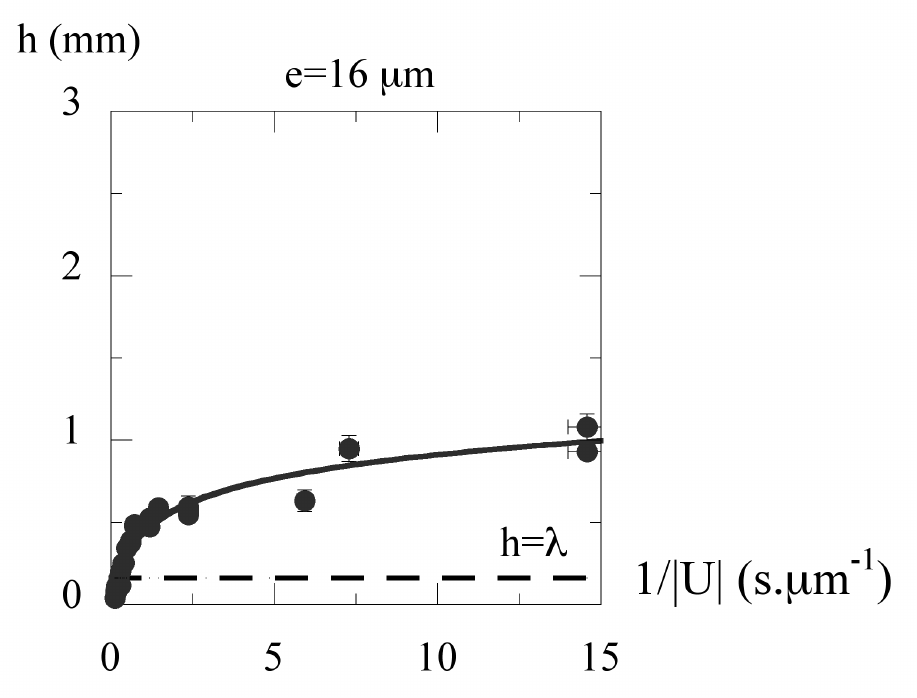}
	\flushleft{\hspace{4cm} (e) \hspace{8cm} (f)}
\caption{
Evolution of the thickness $h$ of the compacted layer with the inverse of the Darcy velocity $|U|$  for various sample depths $e$.
The full lines correspond to the best fits of data to the relation (\ref{h,U,dpdz}) with the critical velocity $U_c$ fixed at $U_c=15 \mu m.s^{-1}$.
The 
\textcolor{black}{dashed}
line shows the Janssen's length $\lambda=10 \; e$ that is identified in figure \ref{G,xi,PT}-a from the fit results.
This length provides the transition between a linear trend with dominant viscous dissipation and a concave trend with a dominant solid friction.
Sample depth : 
(a) $e=125 \mu$m. (b) $e=100 \mu$m.
(c) $e=75 \mu$m. (d) $e=50 \mu$m.
(e) $e=30 \mu$m. (f) $e=16 \mu$m.
The particle diameter is $d=3 \mu$m.
Fluctuations of $h$ and $U$ are indicated as error bars, most of which being smaller than the data symbols.
They have not been taken into account for fitting data.
}
\label{G,h,1-U,NL}
\end{figure*}

\section{Modelling}
\label{Modelling}

The compacted layer ceases to grow when particles begin to be trapped by the solidification front.
The steady thickness that it then reaches is thus linked to the repelling/trapping transition of particles at the front.
We develop below a mechanical model of this transition by considering the forces that act on a particle nearing the front.
In this context, the critical state involved at the repelling/trapping transition will correspond to a net force balance on a particle entering the front.
As some dissipative forces will appear to depend on $h$ and $U$, this balance will provide a link between these variables that will be confronted to our data.

\subsection{Forces on particles}
%

\begin{figure}[t!!!]
\centering
\includegraphics[width=8cm]{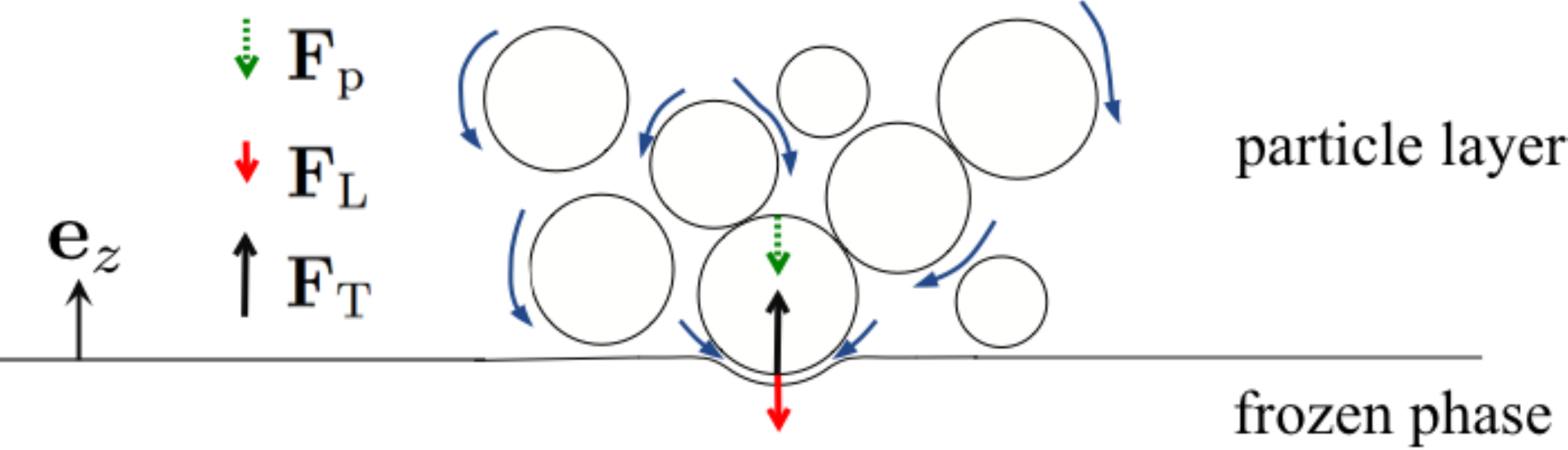}
\caption{(Color online)
Sketch of the forces acting on a particle entering the frozen phase: a thermomolecular repelling force $\mathbf{F}_{\rm T}$ exerted by the solidification front (black arrow); a lubrication force $\mathbf{F}_{\rm L}$ induced by the liquid flowing in the thin premelted film in between the particle and the front (red arrow); a force $\mathbf{F}_{\rm p}$ exerted by the particle matrix on the particle, as a result of the viscous friction and the pressure drop induced by the liquid flow through the whole layer (green 
\textcolor{black}{dotted}
arrow). 
The liquid flow is represented by blue curved arrows. 
The direction $\mathbf{e}_z$, normal to the solidification front, is parallel to the thermal gradient.
Particles have the same diameter and are in contact with each other.
However, their random three-dimensional distribution results in different apparent cross-sections and in an apparent lack of contacts between them.
}
\label{ForcesParticuleEntrante}
\end{figure}

We consider a particle adjacent to a solidification front.
It is surrounded by the front, the suspension fluid and the particle matrix, all of them being suitable to exert a force on it (Fig. \ref{ForcesParticuleEntrante}).
We review these different kinds of forces below :

i) Force exerted by the solidification front  :  thermomolecular force $\mathbf{F}_{\rm T}$ 

It is a short range force that results from van der Waals and electrostatic interactions between the particle and the front.
Its magnitude and intensity varies with materials.
However, it  stands here as a repelling force.

On an elementary particle surface, it yields a normal force whose intensity quickly decreases with its distance to the front, as its inverse cube for non-retarded van der Waals interactions \cite{Israelachvili1992,Wilen1995} and as an exponential for electrostatic interactions \cite{Israelachvili1992,Wilen1995,Wettlaufer1999}.
On a spherical particle, the net force is, by symmetry, parallel to the thermal gradient direction $\mathbf{e}_z$ (Fig.\ref{ForcesParticuleEntrante}). 
Its intensity depends on the distance between the particle base and the front but, as the present particles stand in the critical state of the repelling/trapping transition, we shall assume that this distance and thus the force intensity $F_{\rm T}$ on a particle is a constant.

ii) Force exerted by the fluid :  viscous force $\mathbf{F}_{\rm L}$

 It results from the pressure and viscous stresses exerted by the fluid on the surface of an entering particle.
 It should be emphasized that the thermomolecular force induces an additional pressure between particles and front which, for thermodynamic reasons, maintains a liquid phase between them, whatever the smallness of their distance \cite{Dash2006,Wettlaufer2006}.
 Accordingly, there thus always exists a liquid film, so-called premelted film, in between an entering particle and the front.
 As it is submicronic, it gives rise by far to  the largest dissipation force, $\mathbf{F}_{\rm L}$, that we index by "L" since it corresponds to a lubrication force.
 
This force mainly comes from the depression induced in the premelted film. 
It is thus a trapping force that pushes particles towards the front.
By reason of symmetry, it is parallel to the thermal gradient direction $\mathbf{e}_z$.
As it is induced by creeping flows, its intensity is linearly related to the flow magnitude and thus, to the Darcy velocity $U$ : 
$F_{\rm L} = - f_{\rm L} |U|$, the prefactor $f_{\rm L}$ depending on the geometry of the premelted film (see ref. \cite{Rempel1999,Rempel2001,Park2006} for details) and $F_{\rm L}$ denoting the $z$-component of the force.
In practice, we shall see below that this prefactor is simply related to the critical trapping velocity $U_c$ at which a single particle succeeds to be trapped.

iii) Force exerted by the particle matrix : contact force $\mathbf{F}_{\rm p}$

It is transmitted to a particle nearing the front by contacts between particles along the compacted layer.
It results from the accumulation, along the whole particle matrix, of the stresses induced by the fluid and the sample plates and is thus expected to increase with the layer thickness $h$.
The former stresses refer to the pressure and viscous stresses induced by the fluid flowing across the whole particle matrix. 
The latter stress refers to the solid friction exerted by the sample plates on the particles following the drift of the particle layer pushed by the growing solidification front.

The fluid force is directed in the direction of the Darcy flow $\mathbf{U}$.
The solid friction force opposes the motion of the particle layer with respect to the plates.
It is thus opposite to the growth direction $\mathbf{V}$.
Both forces are thus directed towards the front.
So is their resultant force $\mathbf{F}_{\rm p}$ on an entering particle, which is thus a trapping force. 

As all forces on an entering particle are directed on the $z$-direction, only their magnitude will matter in the sequel.
For convenience, we shall then normalize their $z$-component by the particle cross-section $\pi d^2/4$ to deal with effective mean pressures : the thermomolecular 
\textcolor{black}{pressure}
$\bar{P_{\rm T}}$, the lubrication pressure $\bar{P_{\rm L}}$ and the particle matrix pressure $\bar{P_{\rm p}}$.
The repelling pressure $\bar{P_{\rm T}}$ will then be positive and the two latter trapping pressures, negative.

\subsection{Trapping state}

As the particle layer stands in a critical state for trapping, the trapping pressures $\bar{P_{\rm L}}$ and $\bar{P_{\rm p}}$ have grown up to just reach a force balance on particles nearing the front.
In term of mean pressure, this expresses as : $\bar{P_{\rm p}}+ \bar{P_{\rm L}} +\bar{P_{\rm T}} = 0$.

Here, $\bar{P_{\rm T}}$ is expected to be constant, $\bar{P_{\rm L}}$ is proportional to $|U|$, $\bar{P_{\rm L}}=-g|U|$, and $\bar{P_{\rm p}}$ should depend on the layer thickness $h$.
This offers the opportunity to reduce the number of pressure variables by considering the critical Darcy velocity $U_c$ at which the  lubrication force is sufficient to induce particle trapping on a single isolated particle.
No particle layer can then build up since all particles coming on the front are trapped without delay : $h=0$ and $\bar{P_{\rm p}}=0$.
In this single particle case, the force balance thus reduces to $\bar{P_{\rm L}} +\bar{P_{\rm T}} = 0$ with $\bar{P_{\rm L}} = - g U_c$ and $g= 4 f_{\rm L}/\pi d^2$.
This allows us to express the prefactor $g$ with respect to $\bar{P_{\rm T}}$ and $U_c$, $g=\bar{P_{\rm T}} / U_c$.
In the compacted layer case, this yields at any $|U| \leq U_c$, $\bar{P_{\rm L}} = -  \bar{P_{\rm T}} \; |U|/U_c$, which eventually leads to a relationship between $\bar{P_{\rm p}}$ and  $\bar{P_{\rm T}}$, parametrized by $|U|$ and $U_c$ :
 \begin{equation}
\label{PpPT}
\bar{P_{\rm p}} +  \bar{P_{\rm T}} \; (1- \; |U|/U_c) = 0
\end{equation}
\vskip\baselineskip

To close the 
\textcolor{black}{model}
it now remains to express the mean pressure $\bar{P_{\rm p}} $ with respect to $h$ and $|U|$.
For this, we address below the various stresses involved in the suspension.

\begin{figure}[h!!!]
\centering
\includegraphics[width=8.5cm]{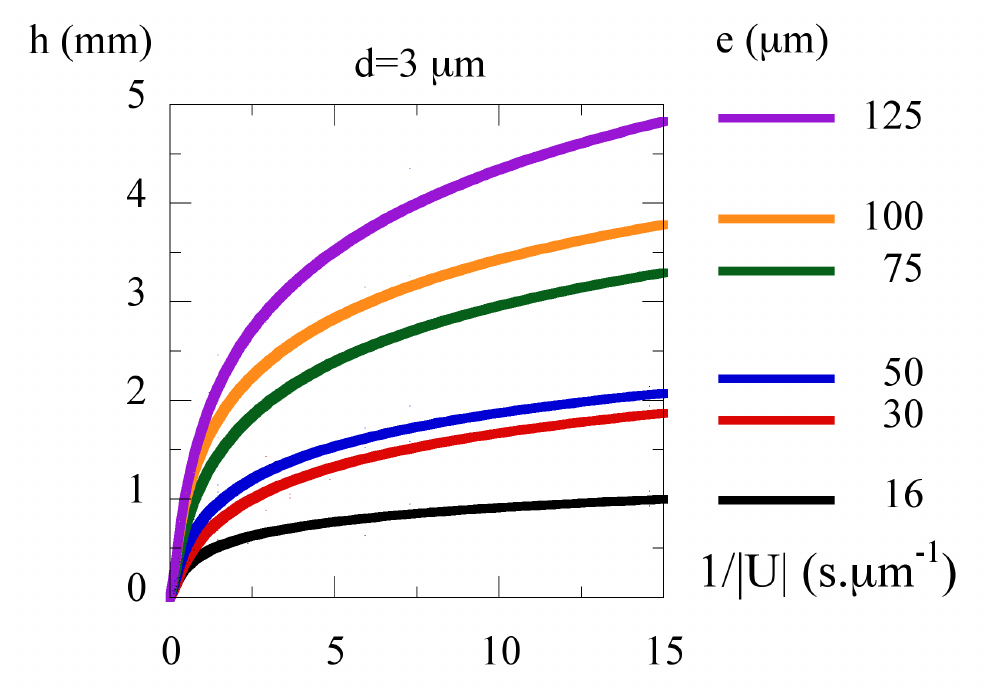}
\caption{(Color online)
Summary of the fits of data to the relation (\ref{h,U,dpdz}) with $U_c$ fixed at $15 \mu$m$.s^{-1}$.
At a given 
\textcolor{black}{At given Darcy velocity, the compacted layer thickness $h$ grows}
with the sample depth $e$.
All show a similar type of concavity, with parameters dependent on $e$.
}
\label{G,h,1_U,Alle,fts}
\end{figure}

\subsection{Suspension stresses}

The mechanical description of a suspension involves the force  experienced by a unit surface of the suspension and the resulting stresses.
As this surface involves both the liquid phase and the particles, part of the force is carried by the contacts between particles while the remaining part is carried by the fluid.
This allows to decompose the suspension stress $\underline{\mathbf{\sigma}}^s$ into a particle stress $\underline{\mathbf{\sigma}}^p$ and a fluid stress $\underline{\mathbf{\sigma}}^f$ : $\underline{\mathbf{\sigma}}^s=\underline{\mathbf{\sigma}}^p + \underline{\mathbf{\sigma}}^f$.

Here the suspension stands in a dense state in which the fluid shear stress is usually neglected in comparison to the frictional particle stress \cite{Andreotti2013}.
In addition, as the compacted particle layer is pushed by the solidification front at a uniform velocity $\mathbf{V}$, it
\textcolor{black}{stands in the front frame in a steady state which can be modeled as a homogeneous Darcy flow.
At the mesoscopic level, this means that the fluid phase experiences no shear stress, so that}
$\underline{\mathbf{\sigma}}^f$ 
\textcolor{black}{reduces to}
the pressure part : 
\textcolor{black}{$\mathbf{\sigma}^{\rm{f}}_{ij}= -p \delta_{ij}$}
where $p$ denotes the fluid pressure.
Steadiness also implies that the suspension stress $\underline{\mathbf{\sigma}}^s$ yields no force on suspension elements : 
\textcolor{black}{$\nabla \underline{\sigma}^s = \mathbf{0}$.}
Altogether, these statements yield the following mechanical equation for the particle phase :
\begin{equation}
\label{NablaSigmap}
\nabla \underline{\sigma}^p = - \nabla \underline{\sigma}^f = \nabla p
\end{equation}

On the other hand, 
the mechanical equation for the fluid phase 
\textcolor{black}{yields the Darcy equation:}
\begin{equation}
\label{nablaP}
\nabla p = - \mu \frac{\mathbf{U}}{k} = \mu \frac{|U|}{k} \mathbf{e}_z
\end{equation}
where the right hand side expresses the volumic viscous friction force in the present porous medium, $k$ denoting its permeability.
In particular, in the present monodisperse suspension, the Kozeny-Carman relation provides the link between the permeability $k$, the particle volume fraction $\phi_l$ of the layer and the particle diameter $d$ :
\begin{equation}
\label{KC}
k(\phi_l,d)=\frac{d^2}{180}  \frac{(1-\phi_l)^3}{\phi_l^2} 
\end{equation}

Relation (\ref{nablaP}) shows that a pressure drop $\Delta p = p(0) -p(h) = - \mu |U| h/k $ sets in the particle layer, between the 
\textcolor{black}{frontier}
 with the suspension ($z=h$) and the solidification front ($z=0$).
It is responsible on one hand for the cryosuction of liquid \cite{Wettlaufer2006} that makes the fluid flow towards the front to feed solidification and, on the other hand, for a part of the trapping force exerted by the particle matrix on 
\textcolor{black}{an}
 entering particle.

It now remains to solve for the particle stress $\underline{\mathbf{\sigma}}^p$ by adding boundary conditions to relation (\ref{NablaSigmap}).

\subsection{Boundary conditions}

We seek to express the boundary conditions imposed to the particle stress $\underline{\mathbf{\sigma}}^p$ at the sample plates by solid friction.
For this, we place the origin of the frame $(x,y,z)$ on the front, at the same distance from the plates and from their lateral boundaries.
Accordingly, the front thus lies in the plane $(x,y,0)$ and the plates in the planes $(x,\pm e/2, z)$.

Following Coulomb's law, solid friction implies a link between the normal stress 
\textcolor{black}{$\sigma^p_{yy}$}
and the tangential stress 
\textcolor{black}{$\sigma^p_{yz}$}
 at the plates :
\begin{equation}
\label{Coulomb}
\sigma^p_{yz} (x,\pm e/2,z)= \mu_W \; \sigma^p_{yy}(x,\pm e/2,z)
\end{equation}
where $\mu_W$ denotes the friction coefficient.

On the other hand, granular materials are known to redistribute stresses on perpendicular directions so that :
\begin{equation}
\label{K}
\sigma^p_{yy}=  K \;  \sigma^p_{zz}
\end{equation}
$K$ designing the Janssen's redirection constant \cite{Janssen1895,JanssenSperl1895}.

Altogether, relations (\ref{Coulomb}) and (\ref{K}) thus yield :
\begin{equation}
\label{BC}
\sigma^p_{yz} (x,\pm e/2,z) =   \mu_W K \;  \sigma^p_{zz} (x,\pm e/2,z)
\end{equation}

\subsection{Janssen effect}
\label{JanssenSubsection}
Considering the $z$-component of relation (\ref{NablaSigmap}) yields the force balance :

\begin{equation}
\nonumber
\partial_x \sigma^p_{xz} + \partial_y \sigma^p_{yz} + \partial_z  \sigma^p_{zz} = \frac{dp}{dz}
\end{equation}

It is convenient to integrate it from $y=-e/2$ to $y=+e/2$ to make connection with the boundary conditions at the sample plates.
Denoting with a tilde the mean variables along the sample depth, one obtains :

\begin{equation}
\label{tildesigma}
 \partial_x \tilde{\sigma}^p_{xz} + \partial_z  \tilde{\sigma}^p_{zz} + \frac{\Delta  \sigma^p_{yz}}{e}= \frac{d\tilde{p}}{dz}
  \end{equation}
where $\Delta  \sigma^p_{yz} = \sigma^p_{yz}(x,e/2,z) - \sigma^p_{yz}(x,-e/2,z)]$.

The symmetry $\sigma^p_{yz}(x,e/2,z)  = - \sigma^p_{yz}(x,-e/2,z)$ and the boundary conditions (\ref{BC}) give :
 \begin{equation}
\nonumber
\Delta  \sigma^p_{yz} =  2 \; \mu_W K \;  \sigma^p_{zz}(x,e/2,z)
  \end{equation}

We may now invoke the following assumptions that are widely used in granular materials \cite{DeGennes1999,Andreotti2013}:
\begin{itemize}

\item a uniform normal stress  $\sigma^p_{zz}(x,y,z)$ in the $y$ direction : 
\textcolor{black}{$\sigma^p_{zz}(x,e/2,z)=\tilde{\sigma}^p_{zz}(x,z)$.}

This assumption, commonly used in granular materials \cite{DeGennes1999,Bertho2003,Ovarlez2003,Andreotti2013}, follows from the propagation of stresses inherent to these materials \cite{Bouchaud1995} and the large uniformity of pressure in piles formed from a uniform rain of grains \cite{Vanel1999}.
In addition, it is supported here by the observations of a flat solidification front, a flat compacted layer and the same layer thickness by both reflection and transmission (Fig. \ref{ReflectionTransmission}).
Following them, the layer thickness $h$ keeps the same value when going from one plate to the other.
As the particle matrix pressure $\bar{P}_p$ depends on $h$ and is linked to the stress $\sigma^p_{zz}$ (see section \ref{ParticleLayerThickness}), the homogeneity of $h$ in between the plates supports that of the particle stress $\sigma^p_{zz}(x,y,z)$ in the $y$ direction.
We finally note that the above considerations refer to stresses, independently of the particle volume fraction.
In particular, the fact that stresses propagate and redistribute in the layer \cite{Bouchaud1995} enables them to homogenize whatever the distribution of particle volume fraction.
 
  \item constant mean stresses $\tilde{\sigma}^p_{ij}$ on the $x$ direction : $\partial_x \tilde{\sigma}^p_{ij} = 0$.
  
  This follows from the Hele-Shaw geometry of the samples for which the sample width $l$ is quite large compared to the sample depth $e$ : $l\gg e$.
  Then, the effect of boundaries on the $x$ direction may be overlooked in most of the suspension so that its mechanical variables may be considered as independent of $x$ in most of the domain.
  This in particular results in :
  
  \item a constant mean tangential stress $\tilde{\sigma}^p_{xz}$ on the $x$ direction : $\partial_x \tilde{\sigma}^p_{xz} = 0$.

\end{itemize}

These assumptions enable us to reduce the stress equation (\ref{tildesigma}) to :
\begin{equation}
\label{equationJanssen}
\partial_z   \tilde{\sigma}^p_{zz} + 2 \; \frac{\mu_W K} {e} \; \tilde{\sigma}^p_{zz} =  \frac{d\tilde{p}}{dz}
\end{equation}
to which we add the boundary condition 
\textcolor{black}{$\tilde{\sigma}^p_{zz}(h) = 0$}
at the 
\textcolor{black}{ frontier}
$z=h$ between the boundary layer and the suspension bulk.
It follows from the fact that particles are not yet into contact there, so that  $\underline{\sigma}^p =\underline{0}$.

As $\phi$ may be considered as independent of $z$  in the compacted particle layer, so must be $d\tilde{p}/dz$ following (\ref{nablaP}) (\ref{KC}) and (\ref{U}).
This enables to easily integrate (\ref{equationJanssen}) to obtain :
\begin{equation}
\label{tildesigmaJanssen}
\nonumber
\tilde{\sigma}^p_{zz}(z) = - \lambda \; \; \frac{d\tilde{p}}{dz} \;\; [ \exp(\frac{h-z}{\lambda}) -1 ]
\end{equation}
with :
\begin{equation}
\label{xi}
\lambda= \frac{e}{2 \mu_W K}
\end{equation}

At the solidification front, $z=0$, this yields the normal stress exerted by the particle matrix :
 \begin{equation}
 \label{sigmatilde}
\tilde{\sigma}^p_{zz}(0) = -  \lambda \; \;  \frac{d\tilde{p}}{dz} \; \; [\exp (h/\lambda) -1 ]
\end{equation}

This exponential response with respect to the height $h$ of the layer is reminiscent of the Janssen effect in granular materials with gravity replaced here by the pressure gradient force $-\nabla p$ \cite{Janssen1895,JanssenSperl1895,Andreotti2013}.
In addition, the characteristic length $\lambda$ of the exponential trend takes the same expression as in the Janssen effect, except that it corresponds here to an amplification length instead of a relaxation length.
This difference traces back to the fact that friction forces are directed here in the same direction as the leading force, the pressure gradient force $-\nabla p$.
Their action thus enhances the effect of the leading force, leading to an exponential 
\textcolor{black}{amplification}
instead of an exponential relaxation otherwise.
The analogous situation in granular  materials would correspond to particles pushed against gravity \cite{Ovarlez2003}.

\subsection{Particle layer thickness}
\label{ParticleLayerThickness}

To determine the particle layer thickness $h$, we need to apply the pressure balance (\ref{PpPT}) on a particle entering the front.
For this, we wish to link the particle matrix pressure $\bar{P}_{\rm p}$ on this particle to the 
\textcolor{black}{mean}
normal stress 
\textcolor{black}{$\tilde{\sigma}^p_{zz}(0)$}
exerted by the particle matrix at the solidification front.

Normal stresses 
\textcolor{black}{$\sigma^p_{zz}(x,y,z)$}
correspond to the $z$-component of the mean force conveyed by particle contacts on a unit surface of the \emph{suspension} normal to the $z$-direction.
However, in average, a part $\varphi$ only of this surface is occupied by particles.
At the level of a particle, this means that the mean force 
\textcolor{black}{$\sigma^p_{zz}(x,y,z)$}
exerted on a unit surface is distributed over particles occupying an average surface $\varphi$.
On each particle, it then corresponds to a mean pressure 
\textcolor{black}{$\bar{P}_{\rm p}(x,y,z)=\sigma^p_{zz}(x,y,z) / \varphi$.}

To determine the particle surfacic fraction $\varphi$, let us notice that it does not depend on $z$ since the compacted layer is homogeneous.
Considering a volume $dV=S dz$, the averaged number of particles in it then writes 
$dN= (\varphi S) dz$.
However, it also amounts to $dN=\phi dV$ by definition of the particle volume fraction.
Accordingly, surfacic and volumic particle fraction appear to be equal : $\varphi= \phi$.

In particular, at the level $z=0$ of the solidification front, the mean pressure on a particle exerted by the particle matrix reads  
\textcolor{black}{$\bar{P}_{\rm p}=\sigma^p_{zz}(x,y,0) / \phi_l$}
and for uniform normal stresses, $\bar{P}_{\rm p}=\tilde{\sigma}^p_{zz}(0) / \phi_l$.
Then, following (\ref{PpPT}) and (\ref{sigmatilde}), one obtains :

\begin{equation}
\label{h,U,dpdz}
h = \lambda \;  \ln[1 +  \; \frac{\chi}{\lambda} \; (\frac{1}{|U|} - \frac{1}{U_c})]
\end{equation}
where
\begin{equation}
\label{chi}
\chi= \phi_l \;  \frac{|U|}{d\tilde{p}/dz} \; \bar{P_{\rm T}}
\end{equation}
is, from (\ref{nablaP}) and (\ref{KC}), independent of $|U|$.

\section{Janssen effect}

We now confront the modelling (\ref{h,U,dpdz}) of the layer thickness $h$ to our experimental data, first to validate the model and the relevance of the Janssen effect to freezing suspension and, second, to extract information on the Janssen's length $\lambda$ and the mean thermomolecular pressure $\bar{P_{\rm T}}$.

For this, we consider the best fit of relation (\ref{h,U,dpdz}) to the data obtained, at each sample depth $e$, at various Darcy velocities $U$ (Fig.\ref{G,h,1-U,NL}).
As the particle diameter $d$ and the fluid viscosity are known, this fit involves three parameters : $(\lambda, \chi, U_c)$.
Among them, the critical velocity $U_c$ involves specific features that need to be discussed.

The critical velocity $U_c$ refers to a force equilibrium on an \emph{isolated} particle entering the front, i.e. for $h=0$.
The forces involved are then the trapping lubrication force at the base of the particle, proportional to $U$, and the repelling thermomolecular force exerted by the front.
As this trapping issue is disconnected from the presence of other particles, it is expected to be independent of $e$.
Microscopic models of trapping \cite{Rempel1999,Rempel2001,Park2006} then provide a determination of $U_c$ that yields $U_c=12 \mu$m.s$^{-1}$ \cite{Saint-Michel2017} in the present context.

However, here, the fits provide $U_c=20 \mu$m.s$^{-1}$ for $e=16 \mu$m and for the other sample depths, either irrelevant negative values or too large positive values of about $30, 80$ or $130 \mu$m.s$^{-1}$.
This means that, above $e=16 \mu$m, the critical velocity is not a sensitive parameter of the fit. 
It is then largely dependent on data fluctuations so that its best fit value is not physically relevant.
As the critical velocity should be a constant of the study, it is then relevant to fix it to a definitive value on the whole data set to prevent its spurious variation with $e$ to affect the fitting of the two other parameters.
In view of the expected theoretical value $U_c=12 \mu$m.s$^{-1}$ and of the observation of vanishing layer thickness close $U=15 \mu$m.s$^{-1}$, we adopt this latter value for the whole data set.
We nevertheless checked that $U_c$ was not essential for the remaining fit parameters $\lambda$ and $\bar{P_{\rm T}}$ so that close values were obtained for them by fixing $U_c$ either to infinity or to $20 \mu$m.s$^{-1}$.

The data fits performed at each sample depth $e$ with fixed parameter $U_c=15 \mu$m.s$^{-1}$ and free parameters $\lambda$ and $\chi$ are displayed in figure \ref{G,h,1-U,NL} and summarized in figure \ref{G,h,1_U,Alle,fts}.
They all agree with 
\textcolor{black}{the}
data and provide values of $\lambda(e)$ and $\bar{P_{\rm T}}(e)$ that we comment below.

\subsection{Influence length $\lambda(e)$}

As shown in figure \ref{G,xi}-a, the influence length $\lambda$ appears to be proportional to the sample depth $e$, $\lambda=10.0 \; e$, in agreement with (\ref{xi}).
This proportionality confirms the relevance of the Janssen effect to the particle layer.
The measured coefficient, $10.0$, corresponds to a product $\mu_W K=0.05$ that we confront below to available data.

The value of the friction coefficient $\mu_W$ depends on materials, here polystyrene for the particles and glass for the sample plates.
It is about $0.35$ for polystyrene-steel and $0.5$ for polystyrene-polystyrene \cite{Tables}.
For polystyrene-glass, although we are not aware of direct measurements of $\mu_W$ for these materials, we may assume its order of magnitude to be $0.5$.
On the other hand, particles are immersed here in the liquid phase of the suspension.
This makes them be surrounded by 
\textcolor{black}{water}
films which are known to decrease friction coefficients by a factor 
\textcolor{black}{at least four between hydrophilic (glass) and hydrophobic (polystyrene) surfaces \cite{Klein2018}.}
Finally, confocal microscopy evidences an ordering of particles at the sample plates (see Fig. \ref{Confocal}) which is known to reduce the friction coefficient by a factor of about $5/6$ \cite{Mandal2017}.
Altogether, one may then expect the effective friction coefficient of particles on the sample plates to be about $0.5 \times (1/4) \times (5/6) \approx 0.1$.

The value of the Janssen's redirection constant $K$ for granular materials stands in between that of solid, $K=0$, and of fluids, $K=1$. 
It amounts to $0.58$ for a compact triangular stack \cite{Durand2010} and is usually taken as about $0.5$ for particles assemblies.

Altogether, we thus obtain an order of magnitude for the product $\mu_W K$ of $0.05$, in agreement with the estimation from fit measurements.
We may compare this value to a measure of $\mu_W K$  obtained from the Janssen effect in the case of glass beads constrained by a polystyrene wall : $\mu_W K\approx 0.37$ \cite{Ovarlez2003}.
Although the present situation is symmetric regarding materials with beads made of polystyrene and walls of glass, the product $\mu_W K$ should be similar except that particles are immersed in 
\textcolor{black}{water}
here.
As this reduces the friction coefficient by a factor 
\textcolor{black}{four at least \cite{Klein2018},}
one may finally estimate $\mu_W K$ to about $0.09$ which is of the order the above estimates.

\begin{figure*}[t!]
\centering
\includegraphics[height=6cm]{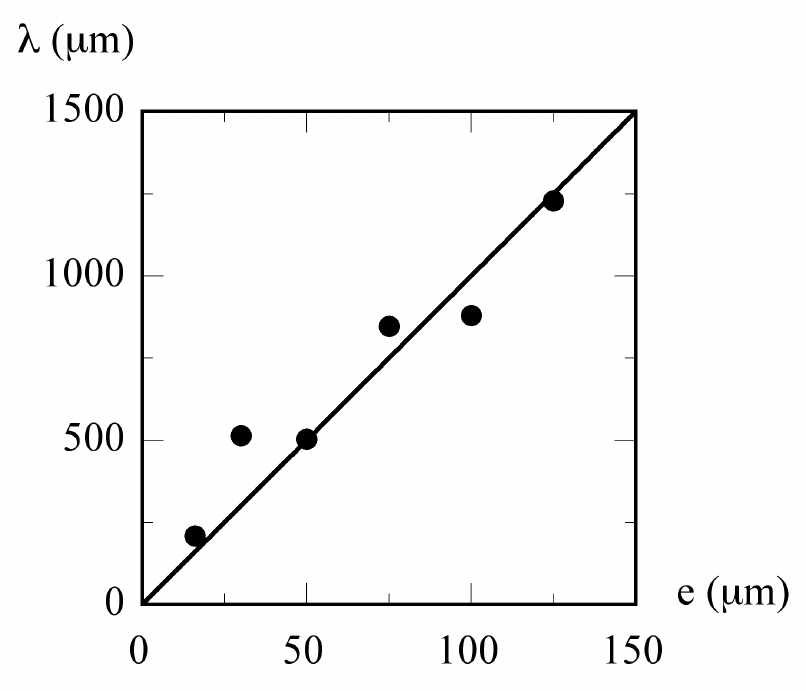}
\hspace{1cm}
\includegraphics[height=6cm]{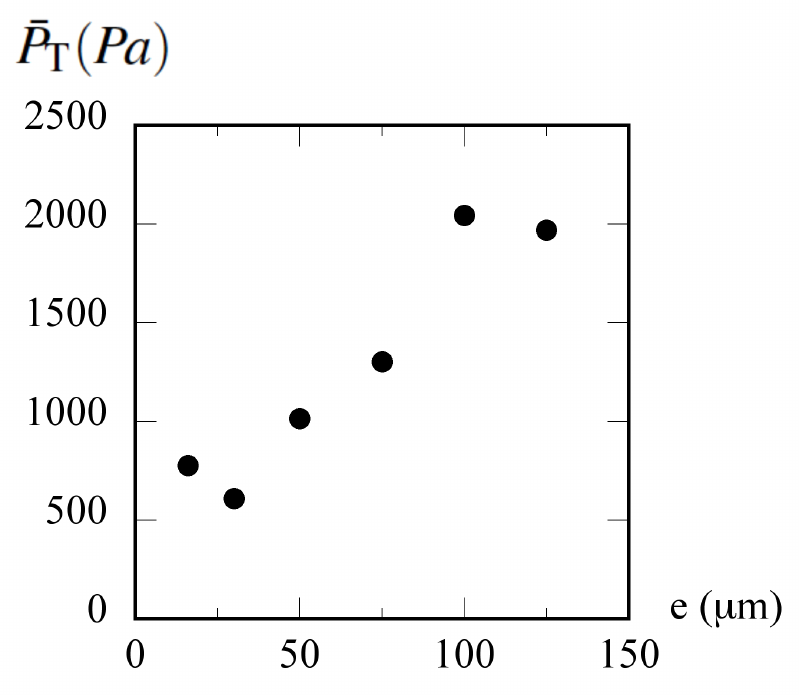}
\\
\flushleft{\hspace{4.5cm} (a) \hspace{7.5cm} (b)}
\caption{
Fit parameters $\lambda$ and $\bar{P_{\rm T}}$ at various sample depths $e$.
They have been obtained from the non-linear evolution of the graphs $h(1/|U|)$ of figure \ref{G,h,1-U,NL}.
(a) Janssen's length $\lambda$ at various sample depth $e$.
The full line shows the best linear fit of data : $\lambda= 10.0 \; e$.
(b) Thermomolecular pressure $\bar{P_{\rm T}}$ at various sample depth $e$.
}
\label{G,xi}
\label{G,PT}
\label{G,xi,PT}
\end{figure*}

\begin{figure*}[t!]
\centering
\includegraphics[height=6cm]{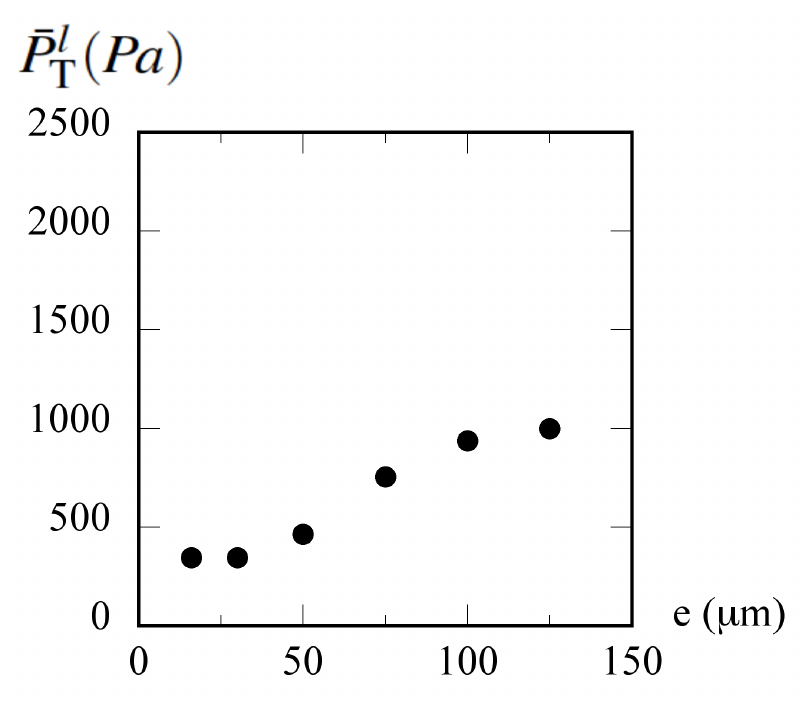}
\hspace{1cm}
\includegraphics[height=6cm]{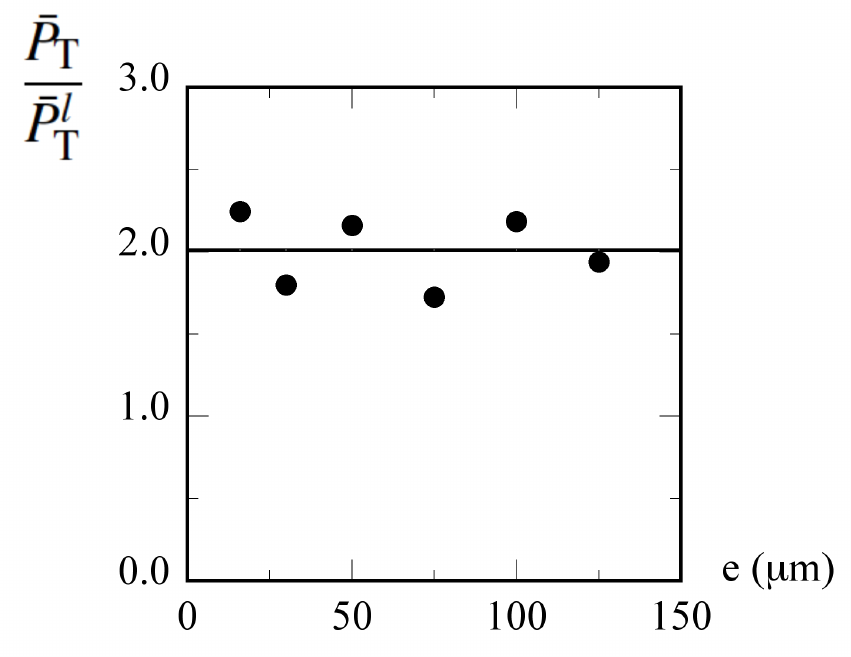}
\\
	\flushleft{\hspace{4.5cm} (a) \hspace{8cm} (b)}
\caption{
(a) Thermomolecular pressure $\bar{P}^l_{\rm T}$ determined from a linear fit of figures \ref{G,h,1-U,NL} in the domain $h < \lambda$.
(b) Ratio of the two values $\bar{P}^l_{\rm T}$  and $\bar{P}_{\rm T}$ of the thermomolecular pressures evaluated from linear fit close to the origin and from non-linear fit on the whole data range. 
The full line shows the average value $2.01$ of the data.
This mean ratio is shown to result from the distance between data and the critical trapping state $h=0$, 
\textcolor{black}{$U=U_c$.}
}
\label{G,PTl,PT-PTl}
\end{figure*}

\subsection{Thermomolecular pressure $\bar{P_{\rm T}}(e)$}

The fit parameter $\chi$ allows from (\ref{nablaP}) (\ref{KC}) and (\ref{chi}) an indirect determination of the mean thermomolecular pressure $\bar{P}_{\rm T}$ over entering particles.
As this pressure corresponds to a local submicronic interaction between a particle and the solidification front, it is expected to be independent of large scale features, in particular the sample depth $e$.
However, figure \ref{G,PT}-b surprisingly displays values that vary nearly over a factor four on our sample depth range.

To address the relevance of the value of $\bar{P}_{\rm T}$ obtained from the fit of the non-linear relation (\ref{h,U,dpdz}) over the whole data range, we restrict attention to the domain $h <\lambda$ which contains most of our data.
Using (\ref{nablaP}) (\ref{KC}) and (\ref{chi}), it then \emph{a priori} refers, at first order in $h/\lambda$, to the linear and $\lambda$-independent relationship : 
\begin{equation}
\label{h,U,phik,ordre1}
h =  \frac{ \phi_l \; k(\phi_l,d)}{\mu}  \;  \bar{P_{\rm T}}\; (\frac{1}{|U|} - \frac{1}{U_c}) + o(h)
\end{equation}
The largest data density and the reduction to a single fit parameter $\bar{P}_{\rm T}$ is then expected to provide a more relevant determination that we label $\bar{P}^l_{\rm T}$.

However, figure \ref{G,PTl,PT-PTl}-a still shows a large variation whereas figure \ref{G,PTl,PT-PTl}-b reveals that both determinations $\bar{P}_{\rm T}$ and $\bar{P}^l_{\rm T}$ are largely correlated since they exhibit a ratio of two : $\bar{P}^l_{\rm T}(e)  \approx \bar{P}_{\rm T}(e)/2$.
This confirms that the variation with the sample depth of the determination of $\bar{P}_{\rm T}$ is independent of the fit method and of the fit domain.
On the other hand, the ratio between the two evaluations is striking since, following the expansion (\ref{h,U,phik,ordre1}), one might have expected equality instead of a doubled value.
To explain it, we question the relevance of this expansion regarding the fit domain $h<\lambda$ of figure \ref{G,h,1-U,NL}.
In particular, it appears that, to provide noticeable values of $h$, most of our data belong to the range $0.3 < h/\lambda <1$ instead of $0 < h/\lambda <1$.
Accordingly, the relevant tangential approximation to relation (\ref{h,U,dpdz}) in our data range should be applied around the center of the actual fit domain, $h/\lambda=0.65$, instead of  $h/\lambda=0$, yielding :
\begin{equation}
\label{h,U,phik,ordre1,y0}
h =  0.17 \lambda +  \frac{ \phi_l \; k(\phi_l,d)}{\mu}  \; \frac{\bar{P_{\rm T}}}{1.9} \; (\frac{1}{|U|} - \frac{1}{U_c}) + o(h-0.65 \lambda)
\end{equation}
In comparison to relation (\ref{h,U,phik,ordre1}), the concavity of relation (\ref{h,U,dpdz}) thus yields the slope of $h(1/|U|)$ to refer to $\bar{P_{\rm T}}$ divided by $1.9$.
This means that the values  $\bar{P}^l_{\rm T}$ determined from (\ref{h,U,phik,ordre1}) were underestimated by a factor $1.9$, in agreement with figure \ref{G,PTl,PT-PTl}-b.

Accordingly, although the range $h<\lambda$ is more suitable for fitting due its largest data density, the resulting values of the thermomolecular pressure will have to be doubled to recover the relevant values.

It nevertheless remains that the thermomolecular pressures determined this way unexpectedly depend on the sample depth $e$.
We stress that this issue is decoupled  from the relevance of relation (\ref{h,U,dpdz}) and thus of the Janssen effect, and only concerns the interpretation of the parameter $\chi$ in terms of  thermomolecular pressure $\bar{P}_{\rm T}$.
We address it in a forthcoming paper \cite{Saint-MichelSubmitted} by questioning the effect of the flatness of the sample plates on the 
\textcolor{black}{particle volume fraction}
 in their vicinity.
Although confocal microscopy reveals a particle ordering at the plates (see figure \ref{Confocal} and ref. \cite{Saint-Michel2017}), the dynamics of the compacted layer build-up shows that its mean particle volume fraction is close to the random close packing value\cite{Saint-Michel2017}.
This suggests that the long-range ordered close packing only holds close to the boundaries, whereas particles in the bulk indeed arrange randomly.
This ordering  echoes the particle ordering observed on few layers close to planar boundaries in monodisperse granular materials \cite{Zhang2006,Burtseva2015,Mandal2017}.
It implies inhomogeneous volume fractions on the sample depth axis with implications on the permeability $k$ from (\ref{KC}) and thus on the viscous stresses undergone by the particle matrix.
Invoking the redistribution of stresses in the particle packing enables us to reevaluate the proportionality between $\chi$ and $\bar{P}_{\rm T}$ and then to reinterpret the variations of slopes near the origin in figures \ref{G,h,1-U,NL} and \ref{G,h,1_U,Alle,fts} as a sole effect of an evolution of volume fraction $\phi_l(y)$ over about $8$ particle diameters and at a single  thermomolecular pressure value \cite{Saint-MichelSubmitted}.

\section{Conclusion}
\label{Conclusion}

For growth velocities below the critical trapping velocity of isolated particles, the solidification of suspensions spontaneously yields an heterogeneous situation in which a dense compacted layer of particles builds up ahead of the solidification front.
As freezing proceeds, this dense layer is pushed by the advancing front and drifts onto the sample boundaries.
Meanwhile, it reaches a steady thickness $h$ that we have measured at various Darcy velocities $U$ and sample depths $e$.
At given $e$, $h$ shows two regimes with an evolution linear in $1/U$ at large $U$ followed, at lower 
\textcolor{black}{$U$,}
 by a much weaker raise.
To understand this evolution, we have linked the steady state of $h$ to a critical trapping state at the front and modeled it by a force balance.
One of these forces, conveyed by the contact between particles,  accumulates all along the particle layer the dissipative effects provided by viscous friction in the layer bulk and 
\textcolor{black}{by}
solid friction at the sample plates.
Expressing the force balance then yields a relationship $h(U,e)$ whose fit to data provides a fine agreement.
This leads the two regimes to be associated to a specific dominant dissipation, viscous friction at large velocities and solid friction at low velocities.

In the solid friction regime, at low velocities, both the logarithmic trend of the curve $h(1/U)$ at each sample depths $e$ and the proportionality of its influence length with $e$ are analogous to the features brought about by the Janssen effect in granular materials.
However, here, gravity is replaced by pressure gradient and friction forces enhance the leading force instead of being opposed to it.
Accordingly, in freezing suspensions, the way friction forces are mobilized appears similar to that involved in granular material.
However, instead of reducing the apparent weight of particle assemblies, the Janssen effect increases here the trapping force applied by the particle matrix on the particles that near the front.
It then yields a macroscopic implication by largely reducing the layer thickness required to make particles enter the advancing solidification front at low growth velocities.

Altogether, these results clarify the formation and the main features of the dense particle layer formed ahead of solidification fronts in freezing suspensions.
In particular, they identify the respective role of the two dissipative phenomena at work in the particle layer, viscous friction and solid friction, and evidence the occurrence of a Janssen effect induced by the latter.
Beyond the model experiment in parallel sample plates performed here, the present insights in the behavior of dense  particle layers could be useful in more natural or complex situations involving largely disperse suspensions with large particles playing the role of boundaries for the thinnest.
Finally, on a more general viewpoint, the evidence of a Janssen effect in this study provides a useful bridge between the issues of freezing suspensions and those of granular materials.

\section*{Conflicts of interest}
There are no conflicts of interest to declare.

\section*{Acknowledgments}

The research leading to these results has been supported by the European Research Council under the European Union's Seventh Framework Program (FP7/2007-2013)/ERC grant agreement 278004.

\bibliography{ReferenceFriction} 
\bibliographystyle{rsc} 

\end{document}